\documentclass{sig-alternate}
\usepackage{tabularx}
\usepackage{subfigure}
\usepackage{paralist}
\usepackage{algorithm}
\usepackage{algorithmicx}
\usepackage{algpseudocode}
\usepackage{multirow}
\usepackage{balance}
\usepackage{tablefootnote}
\usepackage{xspace}
\usepackage[textsize=tiny]{todonotes}

\usepackage{amsthm}

\newcommand{\sys}{Darwini\xspace}
\newcommand{\erdos}{Erd{\"{o}}s-R{\'{e}}nyi\xspace}

\algnewcommand{\LineComment}[1]{\State \(\triangleright\) #1}
\algtext*{EndIf}
\algtext*{EndFor}
\algtext*{EndFunction}

\begin{document}
%

\title{\sys: Generating realistic large-scale social graphs}

%
%
%
%
%

\numberofauthors{5} 
%
\author{
%
%
%
\alignauthor Sergey Edunov\\
       \affaddr{Facebook}\\
       \affaddr{Menlo Park, CA, USA}\\
       \affaddr{edunov@fb.com}
\alignauthor Dionysios Logothetis\\
       \affaddr{Facebook}\\
       \affaddr{Menlo Park, CA, USA}\\
       \affaddr{dionysios@fb.com}
\alignauthor Cheng Wang\\
       \affaddr{University of Houston}\\
       \affaddr{Houston, TX, USA}\\
       \affaddr{cwang35@uh.edu}
\and  
\alignauthor Avery Ching\\
       \affaddr{Facebook}\\
       \affaddr{Menlo Park, CA, USA}\\
       \affaddr{aching@fb.com}
\alignauthor Maja Kabiljo\\
       \affaddr{Facebook}\\
       \affaddr{Menlo Park, CA, USA}\\
       \affaddr{majakabiljo@fb.com}
}

\maketitle
\begin{abstract}
Synthetic graph generators facilitate research in graph algorithms and
processing systems by providing access to data, for instance, graphs resembling
social networks, while circumventing privacy and security concerns.
Nevertheless, their practical value lies in their ability to capture important
metrics of real graphs, such as degree distribution and clustering properties.
Graph generators must also be able to produce such graphs at the scale of
real-world industry graphs, that is, hundreds of billions or trillions of
edges.


In this paper, we propose \emph{\sys}, a graph generator that captures a
number of core characteristics of real graphs. Importantly, given a source
graph, it can reproduce the degree distribution and, unlike existing
approaches, the local clustering coefficient and joint-degree distributions.
Furthermore, \sys maintains metrics such node PageRank, eigenvalues and the K-core
decomposition of a source graph.  Comparing \sys with state-of-the-art
generative models, we show that it can reproduce these characteristics more
accurately.  Finally, we provide an open source implementation of our approach
on the vertex-centric Apache Giraph model that allows us to create synthetic
 graphs with one trillion edges.




\end{abstract}


\section{Introduction}
The availability of realistic large-scale graph datasets is important for the
study of graph algorithms as well as for benchmarking graph processing systems.
Graph processing frameworks such as \cite{Gonzalez:2012:PDG:2387880.2387883,
gonzalez:graphx,Kyrola:2012:GLG:2387880.2387884,DBLP:conf/cidr/WangXDG13} have
been developed to run algorithms on web and social graphs like the ones shown
in Table \ref{tbl:datasets}.  Unfortunately, the applicability of these results
toward industry graphs is limited due to significant differences in both scale
and community structure.  As an example, Twitter reported 320M monthly active
users~\cite{twitter_Q3_2015}, and with an estimated average of 208 followers
per user~\cite{twitter_average:URL}, this is approximately 67B connections.
Facebook has 1.39B active users with more than 400B edges~\cite{ching2015one}.
In 2008, Google found the web graph to contain more than 1 trillion unique URLs
on the web.  It is difficult for these organizations to provide researchers
with access to current industry datasets for a number of reasons.  Shared
datasets must respect user privacy and security
concerns~\cite{Backstrom:2007:WAT:1242572.1242598}. Even when data is public
(e.g. web data), the significant time and resources required to collect and
aggregate this information makes it difficult for most researchers.

\begin{table}[h]
 \center
 \begin{tabular}{| c | c | c |}
  \hline
  \textbf{Graph} & \textbf{\#vertices} & \textbf{\#edges}  \\
  \hline
  \hline
 Yahoo-web~\cite{webscope} & 1.4B & 6.6B \\
  \hline
 UK web graph 2007~\cite{BSVLTAG} & 109M & 3.7B \\ 
  \hline
  Twitter~\cite{Kwak2010} & 40M & 1.5B \\
  \hline
  LiveJournal~\cite{Backstrom2006} & 4.8M & 34M \\
  \hline
  DBLP~\cite{Yang2012a} &  318K &  1M \\
  \hline
 \end{tabular}
 
 \caption{Publicly available graph data sets.}
 \label{tbl:datasets}
\end{table}

Synthetic graph generators provide a way to circumvent these limitations.
Nevertheless, their value lies in the ability to capture important metrics of
real graphs, such as degree distribution, graph diameter and others. For
instance, the accuracy of application simulations depends on the fidelity of
such metrics~\cite{Sala:2010:MGM:1772690.1772778}.  Additionally, since
properties, like degree skew, may even guide the design of graph processing
systems~\cite{Gonzalez:2012:PDG:2387880.2387883}, they must represent realistic
data.  Importantly, graph generators must be able to produce such graphs at
scale since system artifacts or bottlenecks may manifest only on large graphs.
System architects can leverage synthetic graphs for capacity planning by proactively benchmarking
at a scale beyond what is currently available.

While existing graph generation models capture several properties of real
graphs, they fall short in at least one of three important aspects.  First,
they may restrict the model to specific degree distributions. The
Kronecker model~\cite{Leskovec:2010:KGA:1756006.1756039}, one of the most
popular generative models, generates only power-law graphs.  Even though this
is a common model, several real graphs behave differently in
practice~\cite{Sala:2010:BAR:1835698.1835791,
DBLP:journals/corr/abs-1111-4503}. For instance, the Facebook social network
limits the number of friends, invalidating the power-law
property~\cite{DBLP:journals/corr/abs-1111-4503}. In vertex-centric graph
systems, like Pregel~\cite{Malewicz2010} and GraphX~\cite{gonzalez:graphx}, the
degree distribution affects performance by means of the compute and network
load balance.

Second, current approaches do not capture local node clustering properties,
such as the clustering coefficient~\cite{Watts1998} distribution, at a fine
granularity~\cite{Sala:2010:MGM:1772690.1772778,
Leskovec:2010:KGA:1756006.1756039, DBLP:journals/corr/abs-1112-3644}. The BTER
model improves upon Kronecker graphs by allowing non-power law distributions,
but assumes that same-degree nodes also have the same clustering coefficient,
which does not hold in practice ~\cite{Sala:2010:MGM:1772690.1772778}.
Inaccurate clustering coefficient may impact, for instance, the fidelity of
graph partitioning algorithms on the synthetic data.  Consequently, this may
also impact the observed performance of systems that partition input benchmark
graphs prior to processing as an optimization technique~\cite{socialhash}.

Third, existing techniques may not be practical to use. For instance, existing
models may require manual tuning of several parameters.  Alternatively, they
may require model fitting prior to graph generation, which, for large graphs,
incurs high overhead and may not scale~\cite{Sala:2010:MGM:1772690.1772778}.

In this paper, we propose \sys~\footnote{\emph{Caerostris Darwini} is a spider
that weaves one of the largest known webs.}, an algorithm that can generate
graphs with explicitly specified node-degree and clustering coefficient
distributions.  Our algorithm, inspired by the BTER model, constructs graphs in
a block fashion, by interconnecting a scale-free collection of subgraphs.
Unlike current approaches, it does so in a way that allows us to control the
clustering coefficient distribution at a fine granularity.  \sys captures a
number of important metrics observed in real graphs. Notably, and unlike other
methods, it captures the joint-degree distribution of real graphs. 

We provide an open source distributed implementation ~\cite{darwini_jira:URL}
of \sys in the vertex-centric Apache Giraph model.  While the core algorithm is
by design parallelizable and scalable, our ability to generate large graphs is
practically limited by the available computational resources, mainly memory.
However, it is often important to generate graphs beyond the available
capacity, for instance, to perform future capacity planning or to benchmark
disk-backed processing mechanisms~\cite{Roy:2015:CSG:2815400.2815408}.  To
address this challenge, our implementation decomposes graph generation to
multiple tasks by exploiting existing community structure in the original
graph.  The generated subgraphs are subsequently connected based on the
observed structure. 

Our algorithm scales linearly on the size of the output graph. Using our
implementation, we are able to generate synthetic graphs with a trillion
edges in approximately 7 hours on a 200-node compute cluster.  The
\sys implementation is easy to use, requiring as input only the degree
distribution and per degree clustering coefficient distribution of an input
source graph. These distributions can be computed in a scalable manner on very
large graphs, making our approach practical.

This paper makes the following contributions:
\begin{compactitem}

\item We introduce \sys, a graph generating algorithm that can reproduce both
the degree and the clustering coefficient distributions of several real graphs,
including the Facebook social graph with hundreds of billions of edges. To
the best of our knowledge, this is the first algorithm that achieves this
validation.

\item  We provide a distributed implementation of the algorithm on top of the
Apache Giraph model that can generate synthetic graphs with up to one trillion
edges.

\item We provide a thorough evaluation of \sys. First, we show that it can
accurately reproduce the degree and clustering coefficient distributions, as
well as a number of important metrics on different real graphs.  We show that
\sys outperforms existing state-of-the-art graph generation techniques
in terms of accuracy. Second, we benchmark our distributed
implementation and show that it scales linearly on the size of the generated
graph.

\end{compactitem}

The remaining of the paper is structured as follows.  In Section
\ref{sec:algo}, we describe \sys in detail, while in Section~\ref{sec:impl} we
outline the distributed implementation.  Section \ref{sec:eval} contains a
thorough evaluation.  In Section~\ref{sec:related}, we give an overview of
related work. In Section~\ref{sec:concl}, we conclude and discuss future work
in this area.

\section{Algorithm}
\label{sec:algo}
At a high level, \sys receives input as a source graph and generates a
synthetic graph, potentially of a different size, that exhibits similar degree
and clustering coefficient distributions.  The \sys algorithm is split in three
successive stages. In the first stage (Section~\ref{sec:algo_measure}), \sys
analyzes the degree and clustering coefficient distributions of the source
graph and assigns a \emph{target} degree and clustering coefficient to each
vertex of the output synthetic graph, such that it matches the desired
distribution. In the second stage (Section~\ref{sec:algo_connect}), \sys groups
vertices into smaller communities and creates edges within the communities
approximating the target degrees and clustering coefficient.  Finally, in the
third stage (Section~\ref{sec:algo_interconnect}), \sys connects vertices
across communities to match the actual target distributions. In the remaining
of this section, we describe each stage in detail.

\subsection{Assigning degrees and clustering coefficients} 
\label{sec:algo_measure}

In the first stage, \sys assigns a target degree and clustering coefficient to
every vertex of the output graph. Assuming that the desired output graph has
$N$ vertices, we will use $G=(V,E)$ to denote the synthetic output graph, $v_i
\in V, 0\leq i < N$ to denote its vertices, and $(v_i, v_j) \in E, 0\leq i,j <
N$ to denote its edges. Each vertex $v_i$ will have a target degree $d_i$ and a
target clustering coefficient $c_i$.

\sys starts by measuring the degree and clustering coefficient distributions on
the source graph. Specifically, \sys computes (i) $F_{deg}$, the degree
distribution across the entire source graph, (ii) $F_{cc}(d)$, the clustering
coefficient distribution among vertices with degree $d$, for all unique values
of $d$.  Unlike approaches like BTER~\cite{DBLP:journals/corr/abs-1302-6636},
\sys captures the clustering coefficient distribution at a fine granularity. 

Subsequently, for every vertex $v_i \in V$, we first draw $d_i$ from the
$F_{deg}$ distribution. After we have picked $d_i$ for vertex $v_i$, we draw
the target clustering coefficient $c_i$ from the corresponding $F_{cc}(d_i)$
distribution. 

\subsection{Connecting vertices in communities} 
\label{sec:algo_connect}

After calculating the target degrees and clustering coefficients, \sys must add
edges to the vertices in a way that matches these targets. Recall that the
clustering coefficient of vertex $v_i$ in an undirected graph is defined as:
\begin{equation}
\label{eq:cc}
c_i=\frac{2N_{\triangle,i}}{d_i(d_i-1)}
\end{equation}
where $N_{\triangle, i}$ is the number of triangles $v_i$ participates in.
Vertex $v_i$ participates in a triangle with vertices $v_j$ and $v_k$ if
$(v_i,v_j) \in E$, $(v_i,v_k) \in E$ and $(v_j,v_k)\in E$.

Adding edges to match both the degree and target clustering coefficients
directly for each vertex is challenging.  Instead, in this stage \sys first
tries to capture just the number of triangles that each vertex should belong to
in the final output graph.  To understand the intuition behind this, consider
the definition in Equation~\ref{eq:cc} and assume vertex $v_i$ is connected in
such a way that it already participates in the right number of triangles, but
has not yet matched its target degree $d_i$. We can then connect it to other
vertices in a way that does not affect $N_{\triangle,i}$, but helps match
$d_i$.  This way, we are indirectly matching the target clustering coefficient
$c_i$ as well.

\sys adds edges so that each vertex participates in approximately the number of
triangles it should eventually belong to, given its target degree and
clustering coefficient.  To do so, \sys creates smaller communities, or
\emph{buckets}, and connects vertices within the buckets only.  Specifically,
\sys groups vertices according to the number of triangles they must eventually
belong to. 

Consider a bucket with $n$ vertices that we connect randomly according to the
\erdos model. Each edge is included with a probability $P_e$.  Therefore, due
to the independence of edge additions, the probability of any combination of
three vertices in the bucket forming a triangle is $P_{\triangle}=P_e^3$. Since
for each vertex there are $N_{\triangle}=(n-1)(n-2)/2$ possible triangles in
which it can participate, the expected number of triangles for a vertex is:
\begin{equation}
\label{eq:tr_expected}
\hat{N}_{\triangle} = P_{\triangle} \cdot N_{\triangle} = 
P_{e}^3 \frac{(n-1)(n-2)}{2}
\end{equation}

\sys leverages the following two observations.  First, notice from
Equation~\ref{eq:cc} that for all vertices $v_i$ of the graph that participate
in the same number of triangles, the value of the product $c_{e,i}=c_i
d_i(d_i-1)$ is the same. Based on this observation, \sys groups vertices in
buckets according to their $c_{e,i}$ value.  Second, we can construct a bucket
with a desired expected total number of triangles using the \erdos model by
setting the size $n$ of the bucket and the probability $P_e$ appropriately,
based on Equation~\ref{eq:tr_expected}.  Based on this, after adding random
edges according to \erdos with the appropriate value for $P_e$, all vertices in
a bucket will participate in the right number of triangles, in expectation.

Notice that there are different combinations of $n$ and $P_e$ that can achieve
the desired expected number of triangles for a bucket $B$. The choice of the values
must satisfy two conditions. First, a bucket must have enough vertices to
accommodate the expected number of triangles. Assuming that every vertex
participates in the expected number of triangles, that is, $N_{\triangle, i} =
\hat{N}_\triangle$, then from Equations~\ref{eq:cc} and ~\ref{eq:tr_expected}
and since $P_e<1$, we get that:
\begin{equation}
 \label{eq:n_min_result}
 n \geq \sqrt{c_i d_i (d_i - 1)} = n_{B,min}, \forall i \in B
\end{equation}

Second, while in this stage \sys only tries to create the desired number of
triangles, it must still ensure that no vertex significantly exceeds its target
degree, and the wrong choice of $n$ may impact this.
To prevent this from happening, we set $n$ as follows.  Since within a bucket
$B$ with $n$ vertices, any vertex can have at most $n-1$ edges, we require:
\begin{equation}
 \label{eq:n_max_result}
 n \leq \min_{i \in B}(d_i)+1 = n_{B,max}
\end{equation}
This way, we can achieve the desired expected number of triangles without
exceeding the degree of any vertex.

We implement the grouping of vertices in buckets in three successive phases,
described in detail by Algorithms \ref{alg1}, \ref{alg_combine} and \ref{alg2}.
In the following, we explain all the steps, referring back to the detailed
algorithm descriptions where necessary.

\begin{algorithm}
\caption{Group vertices into buckets}
\label{alg1}
\begin{algorithmic}[1]
\State \textbf{Input:} Target degrees $d[i]$, $0 \leq i \leq N-1$
\State \textbf{Input:} Target clustering coefficients $c[i]$, $0 \leq i \leq N-1$

\State $B \leftarrow \{\}$ \Comment{Initialize set of buckets $B$}
\For{$i=0$ to $N-1$}   \label{alg1:start_group}
 \State $c_e \leftarrow c[i] * d[i] * (d[i] - 1)$
 \State bucket $\leftarrow$ selectBucket($B$, $c_e$)  \label{alg1:select_bucket}
 \item[]\Comment{Chooses non-full bucket or adds new bucket in $B$}
 \State bucket.add(i)
 \If{bucket.size > $\min_{j \in bucket}(d[j] + 1)$} \label{alg1:cal_min}
  \State bucket.full()  \label{alg1:bucket_full}
 \EndIf
 \EndFor
\State \Return B
\end{algorithmic}
\end{algorithm}

\textbf{Grouping vertices into buckets.} \sys starts with the
execution of Algorithm~\ref{alg1}. It groups vertices in buckets, based on the
value of $c_{e,i}$, as described above (lines
\ref{alg1:start_group}-\ref{alg1:bucket_full}).  Here, \textit{bucket} is a
data structure that contains a set of vertex indices. We use
\textit{bucket.add(i)} to denote the addition of a vertex $v_i$ and
\textit{bucket.size} to denote the current number of vertices in the bucket. 

As \sys adds vertices one by one to the buckets based on the value of
$c_{e,i}$, more than $n_{B, max}$ vertices may fall in the same bucket. To
handle this, the \textit{selectBucket} procedure (line
\ref{alg1:select_bucket}) searches for a non-full bucket with the same $c_e$ or
allocates a new bucket.  Subsequent vertices with the same $c_e$ are added to
the new bucket.  Note that after \sys adds a vertex to bucket $B$, $n_{B, max}$
is recomputed (line \ref{alg1:cal_min}) to reflect the degree of the newly
added vertex and ensure that a bucket never exceeds the allowed size.  If a
bucket $B$ reaches $n_{B,max}$, \sys labels it as full (lines
\ref{alg1:cal_min}-\ref{alg1:bucket_full}).

\begin{algorithm}
\caption{Merging incomplete buckets}
\label{alg_combine}
\begin{algorithmic}[1]
\State \textbf{Input:} Target degrees $d[i]$, $0 \leq i \leq N-1$
\State \textbf{Input:} Set of buckets $B$ \Comment{Output of Algorithm~\ref{alg1}}
\State $B_u \leftarrow \{ b | b \in B, b.size < n_{b,min}\}$   \label{alg:merge:start}
 \item[]\Comment{Buckets with few vertices}
\State $B \leftarrow B - B_u$     \label{alg:merge:diff}
\State sort($B_u$) \Comment {Sort in order of $c_e$ of each bucket}  \label{algo:combine:sort}
\State bucket $\leftarrow$ emptyBucket()
\State $B$.add(bucket)
\ForAll {$b$ in $B_u$}  \label{algo:combine:loop_start}
 \State bucket.merge($b$)
 \If {bucket.size > $\min_{j \in bucket}(d[j] + 1)$}   \label{algo:combine:test_size}
  \State bucket.full()
  \State bucket $\leftarrow$ emptyBucket()
  \State $B$.add(bucket)  \label{algo:combine:loop_end}
 \EndIf
\EndFor
\State bucket.full()
\State \Return $B$ 
\end{algorithmic}
\end{algorithm}

\textbf{Merging incomplete buckets.} After vertex grouping finishes, some
buckets may not have enough vertices to create the necessary number of
triangles based on \eqref{eq:n_min_result}.  To address this, \sys merges small
buckets into bigger ones. This is implemented in Algorithm~\ref{alg_combine}.
Notice that merging causes vertices with a different value of $c_{e,i}$ to be
placed in the same bucket.  As a result there is no single value for $P_e$ that
will approximate well $N_\triangle$ for all vertices in a merged bucket.
Eventually, this may prevent vertices from approximating well the target
clustering coefficient.  Nevertheless, we have found empirically that this
offsets the inaccuracy caused by incomplete buckets.

Besides, \sys merges buckets in a way that mitigates this effect.  After
obtaining all incomplete buckets (lines
\ref{alg:merge:start}-\ref{alg:merge:diff}), it orders them according to their
$c_e$ value (line \ref{algo:combine:sort}). Subsequently, it merges buckets
with close values (lines
\ref{algo:combine:loop_start}-\ref{algo:combine:loop_end}). When it creates a
merged bucket with the maximum allowed size, it marks it as full and allocates
a new one (lines \ref{algo:combine:test_size}-\ref{algo:combine:loop_end}).
This ensures that the expected number of triangles for each vertex in a
bucket is closer than in a random assignment.

\begin{algorithm}
\caption{Create random edges within buckets }
\label{alg2}
\begin{algorithmic}[1]
\State \textbf{Input:} Target degrees $d[i]$, $0 \leq i \leq N-1$
\State \textbf{Input:} Target clustering coefficients $c[i]$, $0 \leq i \leq N-1$
\State \textbf{Input:} Set of buckets $B$ \Comment{Output of Algorithm~\ref{alg_combine}}
\For{$b \in B$}
\State $k \leftarrow arg\min_{k \in b}d[k]$
\State $n \leftarrow b.size()$
\State $p \leftarrow \sqrt[3]{c[k]} $
\For{$i \in b$}
\For{$j \in b$, $j < i$}
\If{random() < $p$}
\State $addEdge(v_i, v_j)$
\EndIf
\EndFor
\EndFor
\EndFor
\end{algorithmic}
\end{algorithm}

\textbf{Adding edges.} After grouping the vertices into buckets, \sys adds
random edges in each bucket according to the \erdos model, to create the
expected number of triangles in the bucket.  Algorithm~\ref{alg2} describes
this process. 

\sys picks the edge probability $P_e$ based on Equations \ref{eq:cc} and
\ref{eq:tr_expected}: 
\begin{equation}
\label{eq:prob}
P_e = \sqrt[3]{\frac{c_i d_i(d_i-1)}{(n-1)(n-2)}}
\end{equation}
Recall that for each bucket $B$, we set the size of the bucket to $n=\min_{i
\in B}(d_i+1)$. We also know the value of the product $c_i d_i (d_i-1)$. Since
the product is similar for all vertices in the bucket, we can pick $d_i$ and
$c_i$ for the vertex with the minimum degree in the bucket. Replacing this in
Equation~\ref{eq:prob} gives $P_e=\sqrt[3]{c_k}$, where $k=arg\min_{i \in
B}d_i$, for bucket $B$.

At the end of this stage, \sys has created the expected number of triangles in
each bucket, but not the target degrees and clustering coefficient. In fact,
for every vertex, its degree should be less than the target degree, therefore
the clustering coefficient should be higher than the target. In the following
section, we describe how \sys correct this.

\subsection{Interconnecting communities}
\label{sec:algo_interconnect}

The previous step created vertices each with degree $d_i'$, smaller than the
target degree $d_i$. In this step, \sys attempts to add the residual degree
$d_i-d_i'$ for each vertex while leaving the number of triangles a vertex
participates intact. This way, it indirectly meets the target clustering
coefficient of a vertex as well.

\sys achieves this by connecting vertices that belong to different buckets,
picking randomly from the entire graph. Intuitively, this increases the
degree of each vertex, but, since the connections are now random across the
entire graph, they are unlikely to contribute to the number of triangles for a
vertex. 

\sys implements this stage with Algorithms \ref{alg3} and \ref{alg4}. In
particular, \sys executes these algorithm iteratively in an alternating manner.

\begin{algorithm}[h]
\caption{Create random edges across buckets}
\label{alg3}
\begin{algorithmic}[1]
\State \textbf{Input:} $V \leftarrow $ Vertices with current degrees $d_{curr}[i]$ 
\State \textbf{Input:} Target degrees $d[i]$, $0 \leq i \leq N-1$
\For{$v_i \in V$}  \label{algo:connect:loop_start}
\If{$d_{curr}[i] < d[i]$}  \label{algo:connect:deg_test}
\State $v_j \leftarrow random(V)$  \Comment{Returns random vertex} \label{algo:connect:rand}
\If{$d_{curr}[j] < d[j]$}  \label{algo:connect:dest_deg_test}
\State $addEdge(v_i, v_j)$  \label{algo:connect:add_edge}
\EndIf
\EndIf
\EndFor
\end{algorithmic}
\end{algorithm}

In every iteration, Algorithm~\ref{alg3} makes a pass on every vertex
(line~\ref{algo:connect:loop_start}).  If a vertex has not met its target
degree yet (line~\ref{algo:connect:deg_test}), it randomly picks a candidate
vertex to connect to (line~\ref{algo:connect:rand}). If by connecting to the
candidate vertex we do not exceed the target degree of the candidate
(line~\ref{algo:connect:dest_deg_test}), then \sys adds an edge between the two
vertices (line~\ref{algo:connect:add_edge}).

\textbf{Satisfying high degree vertices.} During this process, \sys can
easily find candidate edges to satisfy the target degree for the low-degree
vertices. However, as \sys adds edges, it becomes increasingly hard to find
candidates to connect high-degree vertices. This problem manifests in BTER as
well, a problem reported in \cite{DBLP:journals/corr/abs-1302-6636}, and
something that we show in our evaluation as well. 

This problem appears because of the random selection of candidates vertices.
At the same time, for scalability purposes, we want to avoid searching the
entire set of vertices for candidates. To address this, \sys randomly shuffles
vertices into groups and searches for candidates only within the groups. After
each iteration, the size of the groups increases exponentially, gradually
increasing the search space. Algorithm~\ref{alg4} implements this logic.

\begin{algorithm}[h]
\caption{Create edges for high-degree nodes}
\label{alg4}
\begin{algorithmic}[1]
\State \textbf{Input:} $V \leftarrow $ set of $N$ vertices with target degrees  $d[i]$
\State \textbf{Input:} $iter \leftarrow $ current iteration
\State $n_{group} = N/2^{iter}$   \Comment{Group size}
\State $G \leftarrow$ shuffle($V$, $n_{group}$)  \label{algo:group:shuffle}
\item[]\Comment{Shuffles vertices in groups of size $n_{group}$}
\For{$g \in G$}
\For{$v_i, v_j \in g, i<j$}
 \State $p = \frac{|d[i] - d[j]|}{d[i] + d[j]}$
 \If{random() > $p$} \label{algo:coll:rand}
  \State $addEdge(v_i, v_j)$
 \EndIf
\EndFor
\EndFor
\end{algorithmic}
\end{algorithm}
Note that the random shuffling (line \ref{algo:group:shuffle} helps ensure that
\sys does not increase the number of triangles in the graph. More specifically,
the shuffling procedure finds those vertices that have not still met their
target degree randomly partitions them to a set of groups of a specified size.
Within such group, every pair of vertices is a candidate for adding an edge.

\textbf{Maintaining the joint-degree distribution.} Aside from the clustering
coefficient, \sys also attempts to produce a realistic joint-degree
distribution.  \sys is based on the observation that in social networks, there
is a positive correlation between the degree of a node and the degrees of the
neighbors of the node~\cite{DBLP:journals/corr/abs-1111-4503}.

\sys enforces this by randomizing the edge creation process and ensuring that
the probability of creating an edge between vertices with similar degrees is
higher than the probability of creating an edge between vertices with very
different degrees (line~\ref{algo:coll:rand}). As we show in
Section~\ref{sec:eval}, this helps maintain an accurate joint-degree
distribution.

Algorithm~\ref{alg4} ensures this by adjusting the probability of an edge
creation depending on how similar the degrees of the two candidate vertices
are. \sys sets this probability to be equal to $(|d[i] - d[j]|)/(d[i] + d[j])$.
While there are different ways to set the probability, we have found that this
works well in practice.

\section{Implementation} 
\label{sec:impl}
We have implemented \sys on top of Apache Giraph~\cite{ching2015one}
vertex-centric programming model. Here, we give an outline of the
implementation of each algorithm described in Section~\ref{sec:algo}. The
implementation is available as open source at \cite{darwini_jira:URL}.

\subsection{Graph generation}

Using the vertex-centric abstraction, we map each vertex of the output graph to
a Giraph vertex. Our implementation begins by generating the desired number of
vertices on the fly. Vertices are in-memory objects distributed across our
compute cluster, and they contain (i) the IDs of their neighbors and (ii)
computational state specific to the \sys algorithm, that is, the bucket in
which it belongs and its target degree and clustering coefficient.  \sys
initializes the vertices by assigning to each vertex the target degree and
clustering coefficient, drawn from the distributions computed in the first
stage.

Next, we must assign vertices to buckets as per Algorithm~\ref{alg1}. We
experimented with two different implementations of Algorithm~\ref{alg1}. Our
initial implementation leverages the Giraph \emph{master computation},
executing the logic centrally for the entire graph. The master computation,
calculates a vertex-to-bucket assignment for each bucket and broadcasts it to
all worker machines. This way, every vertex picks up their bucket assignment
and saves it in its state.  We also evaluated a parallel implementation, where
each machine in the cluster is responsible for a portion of all vertices and
runs the same algorithm locally.  However, we did not observe any significant
difference in the quality of generated graphs for large graphs. For small
graphs, it is always possible to use the centralized approach.

For Algorithm~\ref{alg2}, we must implement the random edge creation within a
bucket as a per vertex computation. Notice that to implement this logic, we
need information from all vertices in the bucket. After the execution of
Algorithm \ref{alg1}, each vertex knows which bucket it belongs too. For each
bucket \sys picks a bucket representative vertex. To do this, vertices assigned
to the same bucket coordinate with each other and elect as representative the
vertex with the smallest vertex ID.  After that, each vertex sends its target
degree and clustering coefficient along with its own ID in a message to
their bucket representative vertex. After receiving these values, the
representative vertex now has all the information necessary to implement the
logic of Algorithm~\ref{alg2}. After it decides which edges to add, it sends
an edge addition request to the corresponding destination vertices. 

In Algorithm~\ref{alg3}, we implement the random destination vertex selection
as another vertex computation. Unlike the implementation of
Algorithm~\ref{alg2}, a vertex can now pick a destination across the entire
graph. In the \sys implementation, each vertex sends an \emph{edge request
message} to a random destination vertex ID in the graph.  Since the range of
IDs is known, vertices pick one uniformly. Once the destination vertex receives
the request message, if it has residual node degree, it can accept the request.
It adds the edge locally and sends an \emph{edge confirmation message} back to
the sending vertex. At this point, the sending vertex can also add this edge. 

Algorithm~\ref{alg4} is intended to find connections for high degree vertices.
Here, we use the same idea of communities and representative vertices as with
the implementation of Algorithm~\ref{alg2}. Representative vertices will now
correspond to the groups calculated in Algorithm~\ref{alg4}. Here, in each
iteration, every vertex picks a random representative vertex and sends its
target degree and current degree.  The logic for selecting a representative
vertex is similar to that of Algorithm~\ref{alg2}.

\subsection{Scaling beyond cluster capability.} While our implementation is
parallelizable, our ability to generate large graphs is limited by the
available main memory memory. This may be sufficient for medium size graphs up
to billion vertices and trillion edges. However, our goal is to be able to
generate graphs bigger than what our current infrastructure can hold.

To address this, we leverage the observation that in real social networks,
users typically belong in large communities that are relatively sparsely
connected with each other.j. Communities defined by the user country of origin
is such an example. For instance, it has been estimated in
\cite{DBLP:journals/corr/abs-1111-4503} that 84\% of the total number of edges
are within the communities defined by the user country.  These communities
contain a number of vertices that is much bigger than what makes a bucket in
\sys; they may contain hundreds of millions of vertices.  We call these large
vertex groupings \emph{super-communities}. 

Once these super-communities are identified, we first run \sys on each
super-community individually, generating the corresponding synthetic graph.
After this, each synthetic super-community approximates the degree and
clustering coefficient distributions of the original only. We can break this
task into multiple independent ones that require only one super-community at a
time to fit in the available memory.

Next, we need a way to connect vertices across the super-communities. As with
connecting vertices across buckets, we can still connect edges in a random
fashion.  However, we must implement this in a way that does not require
loading the entire graph in memory. Notice that to construct these edges, we do
not need to load the graph structure of each super-community, that is, the
edges of each vertex.  We only need to load the super-community that each
vertex belongs to and its residual degree.  From then on, we essentially repeat
Algorithms \ref{alg3} and \ref{alg4}. This reduces the required amount of
memory by orders of magnitude, allowing us to generate graphs with several
trillions of edges.

\section{Evaluation} 
\label{sec:eval}
\begin{figure*}[t]
  \centering
  \subfigure[Degree distribution] {
   \includegraphics[width=0.31\linewidth]{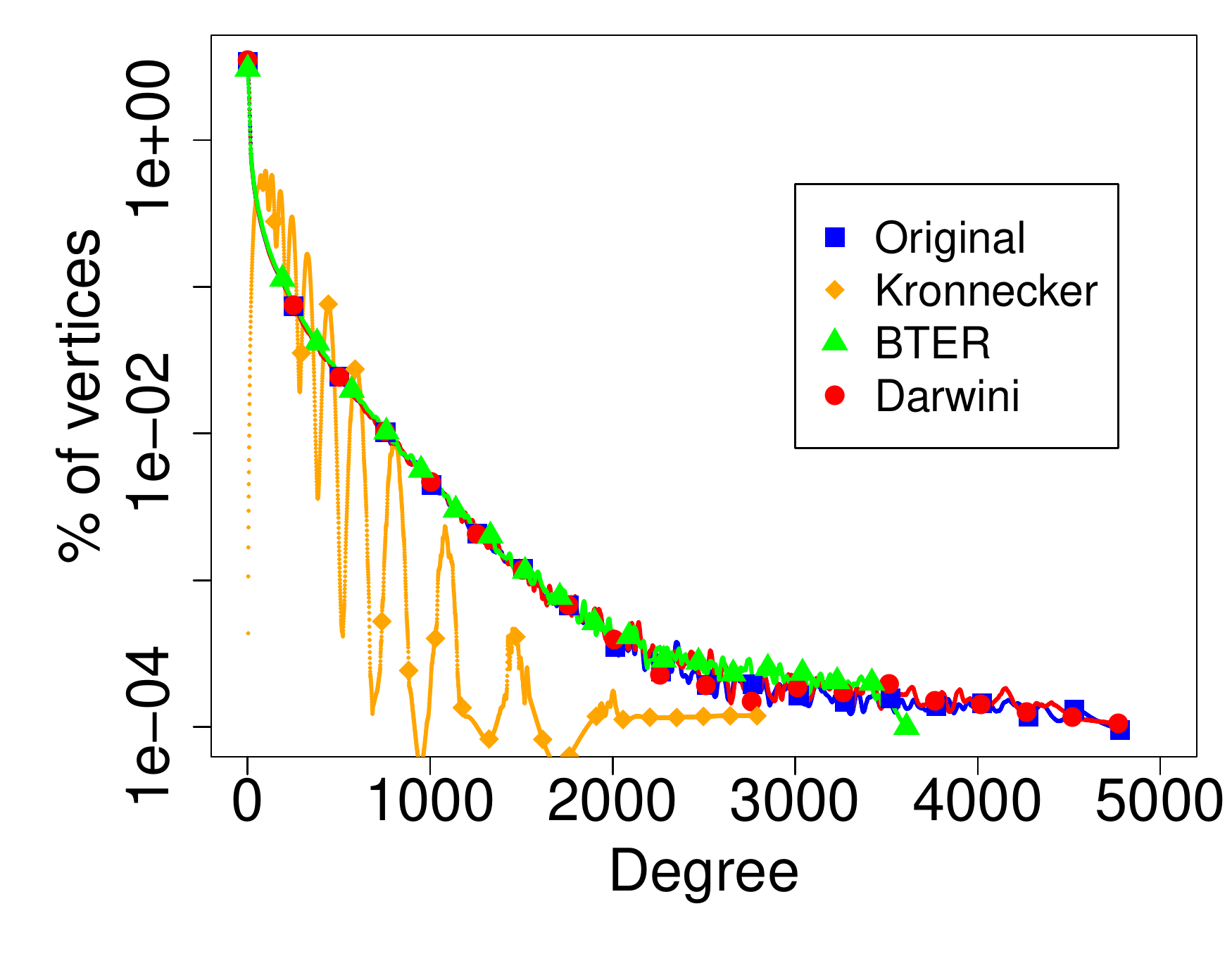}
   \label{fig:ndd}
  }
  \subfigure[Average clustering coefficient] {
   \includegraphics[width=0.31\linewidth]{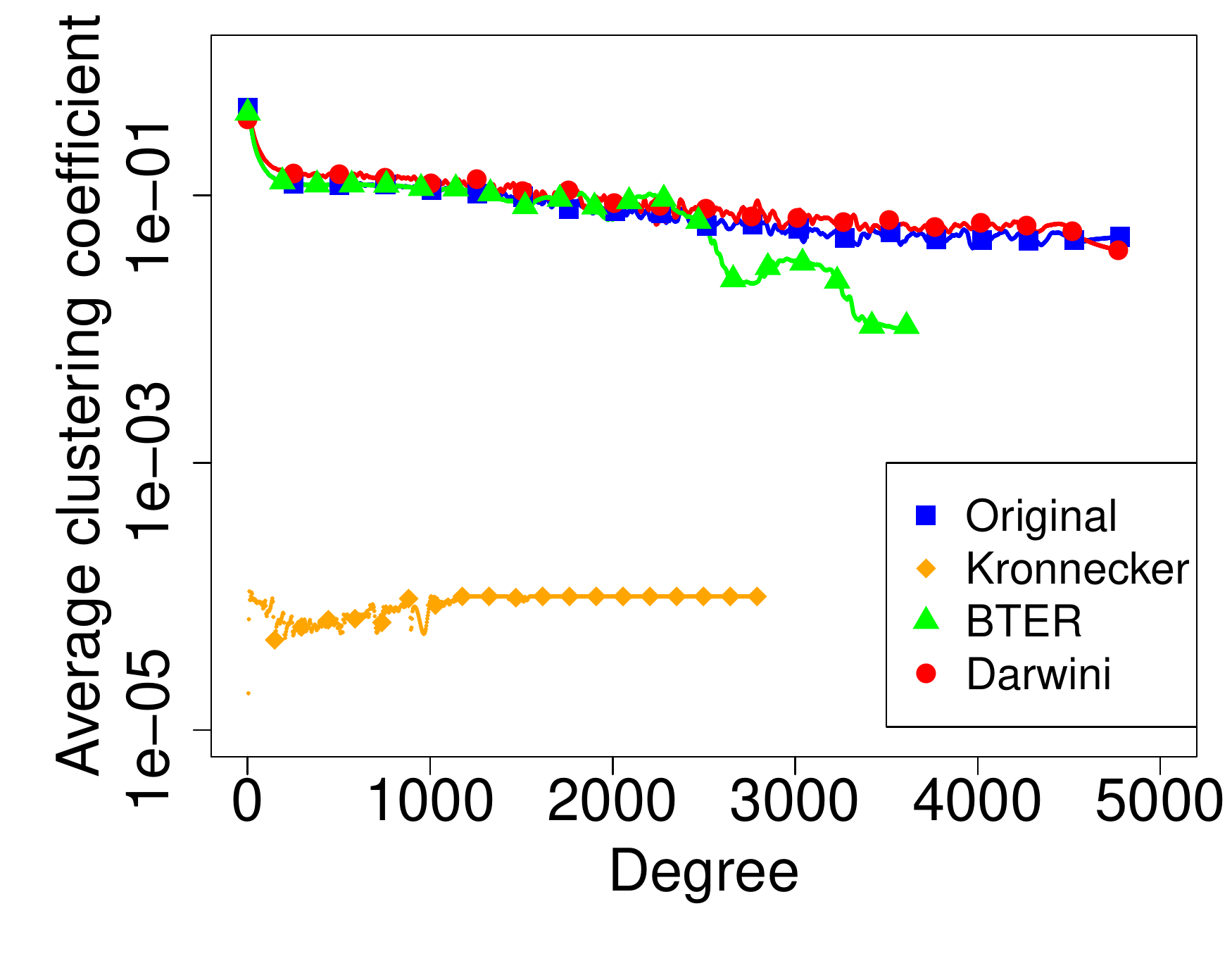}
   \label{fig:avg_ccd}
  }
  \subfigure[Clustering coefficient] {
   \includegraphics[width=0.31\linewidth]{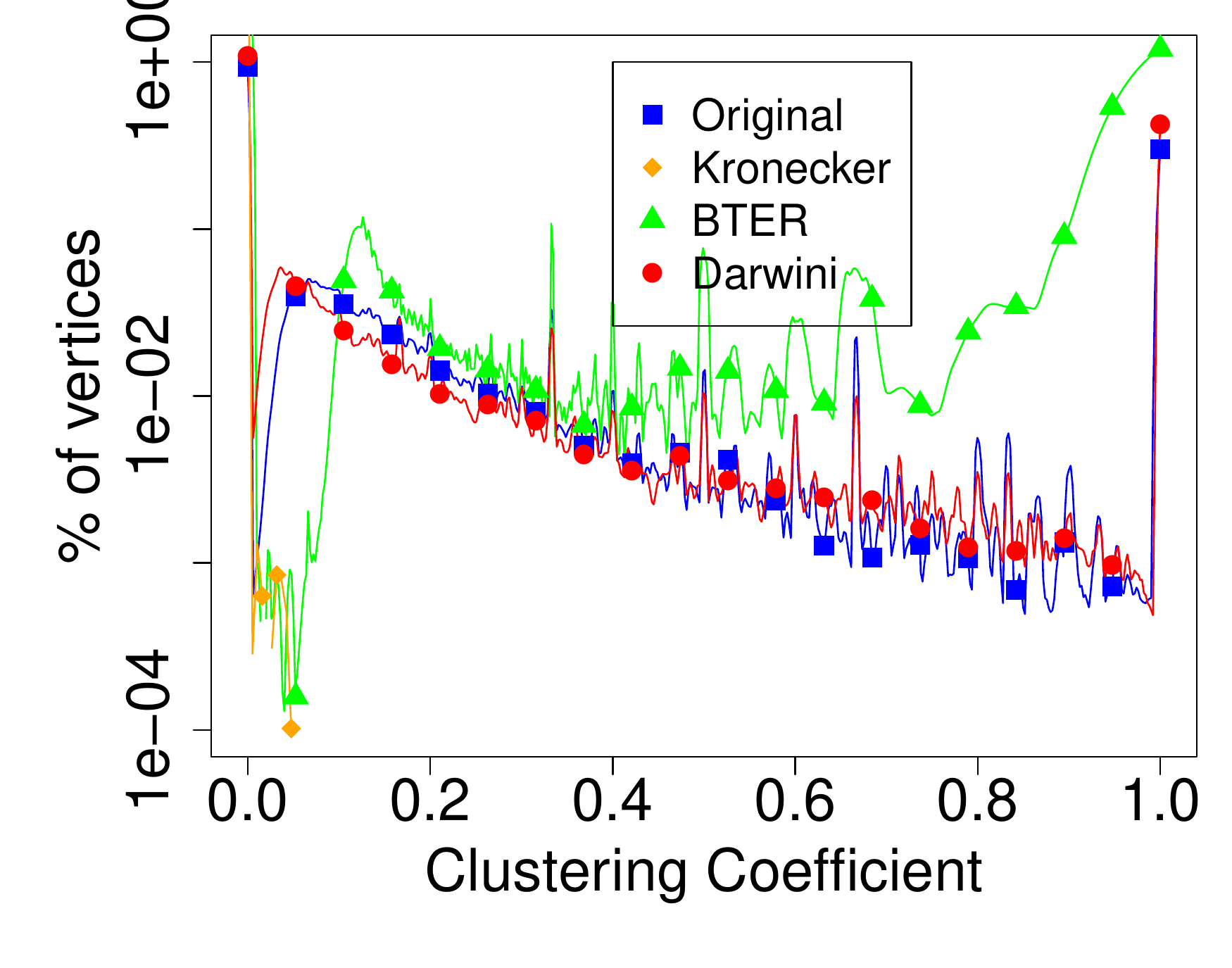}
   \label{fig:ccd}
  }
  \caption{Comparing \sys with Kronecker and BTER under different graph metrics
on the Facebook subgraph. \sys outperforms the other techniques in all metrics.} 
  \label{fig:distr}
\end{figure*}

In this section, we evaluate different aspects of our
algorithm. First, we measure the ability of the algorithm to
accurately capture a number of important graph metrics, and compare
our approach with state-of-the-art generative models. Second, we
measure the impact of this accuracy on application-defined
metrics. Finally, we evaluate the scalability of the algorithm and
measure the computational overhead of our implementation.  

\subsection{Graph metrics}

We start by measuring how accurately our algorithm re-produces a number of
graph metrics, compared with the input source graph. There is a variety of
metrics used to characterize graphs, here we focus on degree distribution,
local clustering coefficient, joint-degree distribution as they directly
characterize the structure of a graph. We also measure the PageRank
distribution, Eigenvalues, K-Core decomposition, and Connected Components as
higher-level metrics.

\subsubsection{Degree distribution}

Here, we measure how accurately \sys reproduces the  degree distribution,
compared with other techniques. We first evaluate the algorithm using a portion
of the Facebook social network as the source graph.  Specifically, we use a
subgraph of the Facebook social graph that represents a specific geographic
region with approximately 3 million vertices and 700 million
edges~\footnote{For confidentiality reasons we cannot provide more information
on the graph.}.  Here, we compare \sys with the BTER and Kronecker models as
they are the only models we could evaluate for a graph of this size. 

In Figure~\ref{fig:ndd}, we compare the degree distribution achieved by the
different models with that of the original graph. First notice, that the
Kronecker model fails to re-produce the degree distribution, as the Facebook
graph does not follow the power-law model.  Even though BTER provides a better
approximation of the degree distribution than Kronecker, it fails to create
high-degree vertices. As the algorithm tries to connect high-degree nodes to
achieve the right clustering coefficient, it fails to find enough candidates.
\sys, instead, produces a degree distribution that is close to the original for
all values of node degree.

\begin{figure}[ht]
  \centering
  \subfigure[Degree] {
   \includegraphics[width=0.48\linewidth]{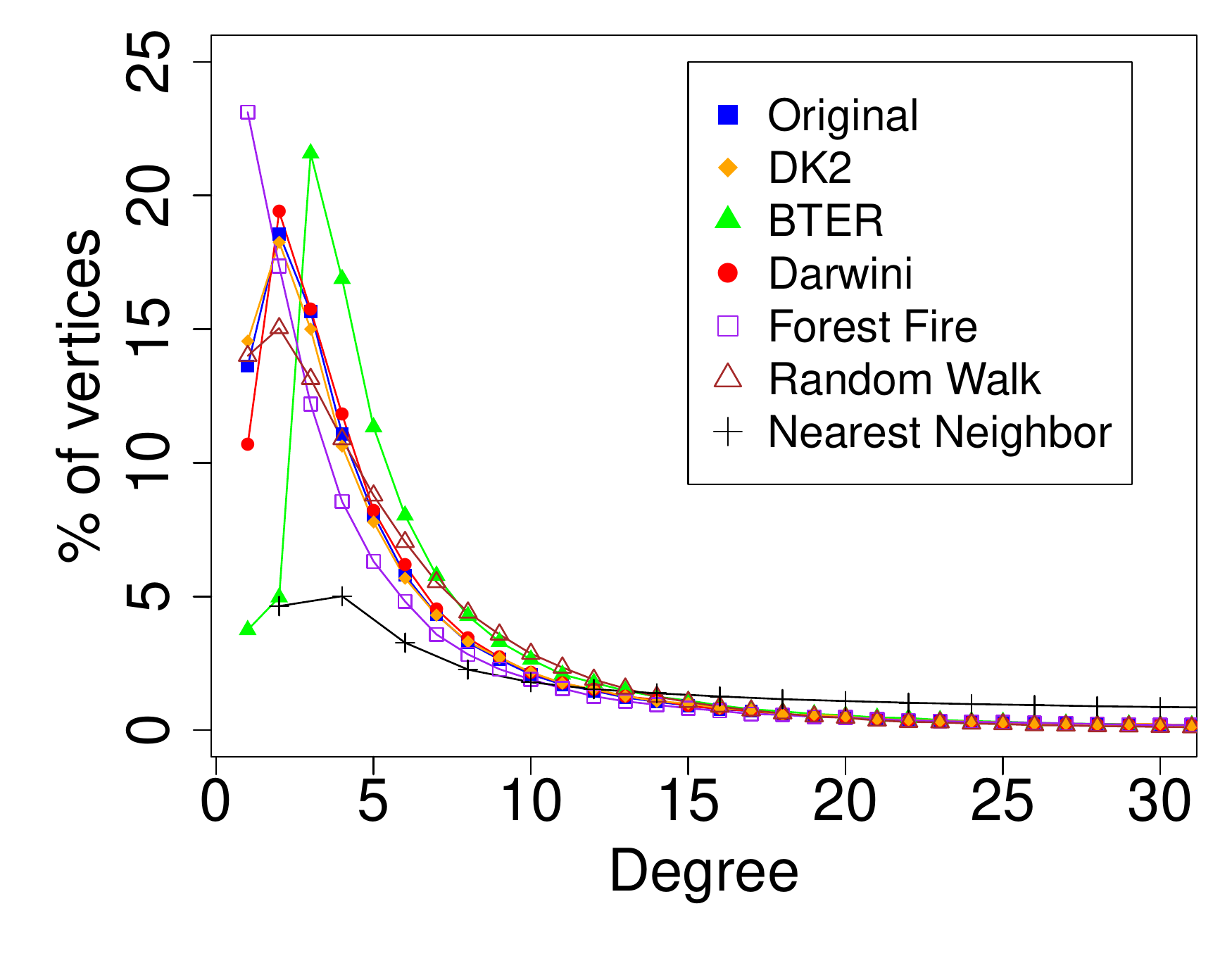}
   \label{fig:dblp_degree}
  } 
  \hspace{-1.7em}
  \subfigure[Avg. clustering coefficient] {
   \includegraphics[width=0.48\linewidth]{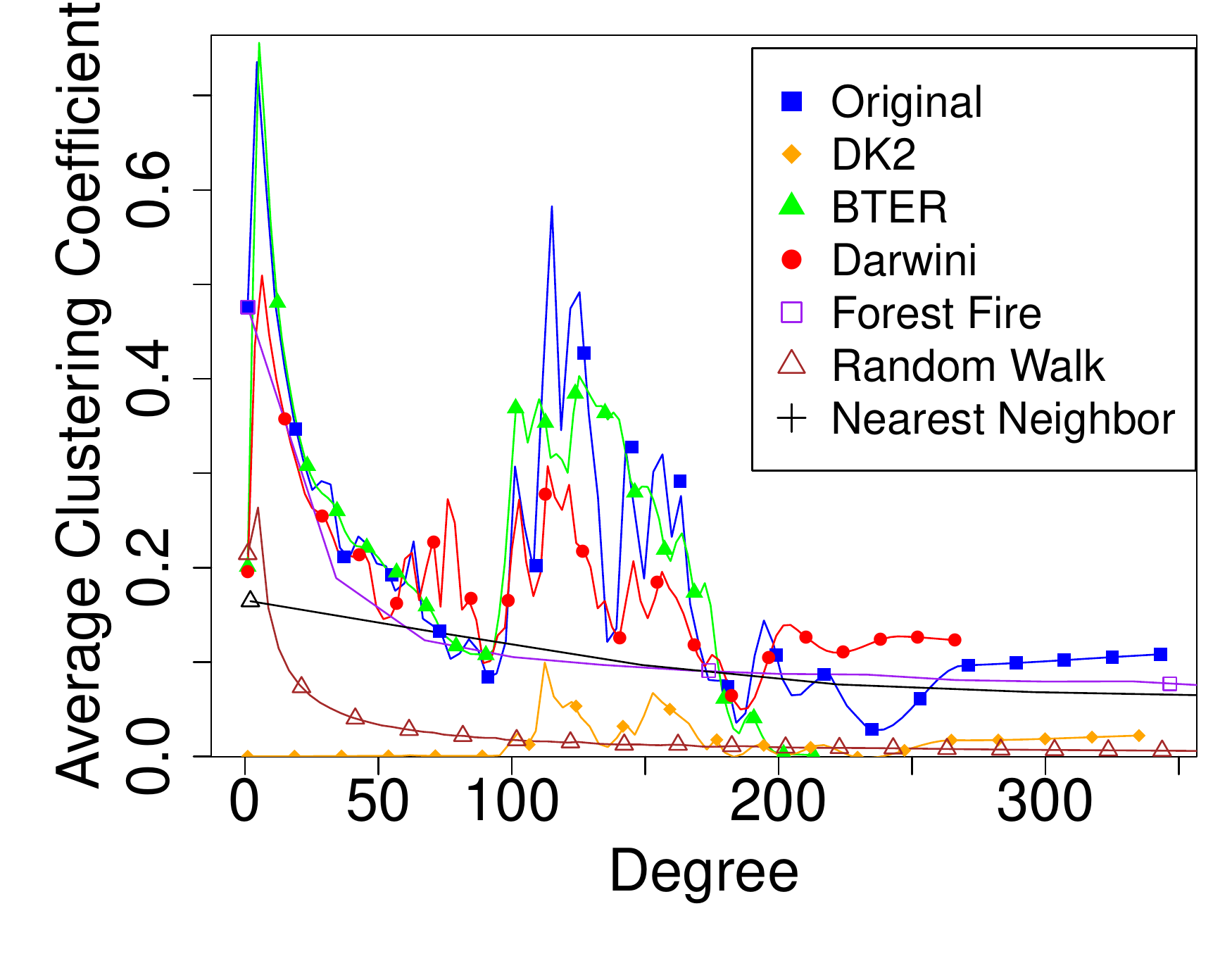}
   \label{fig:dblp_cc}
  }
  \caption{Comparison with several models on the DBLP graph.}
   \label{fig:dblp}
\end{figure}

\begin{figure}[ht]
  \centering
  \subfigure[Degree] {
   \includegraphics[width=0.48\linewidth]{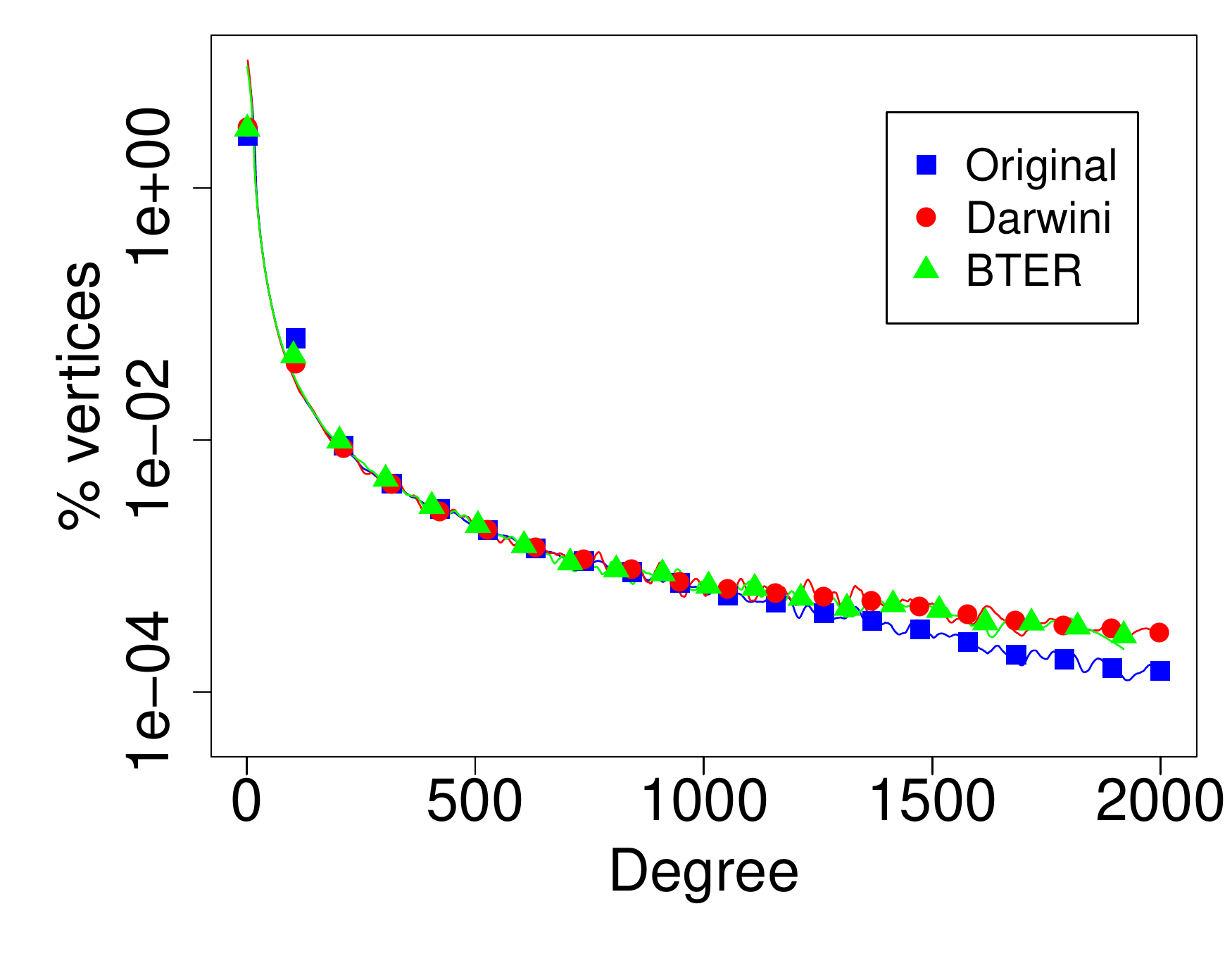}
   \label{fig:twitter_degree}
  }
  \hspace{-1.7em}
  \subfigure[Avg. clustering coefficient] {
   \includegraphics[width=0.48\linewidth]{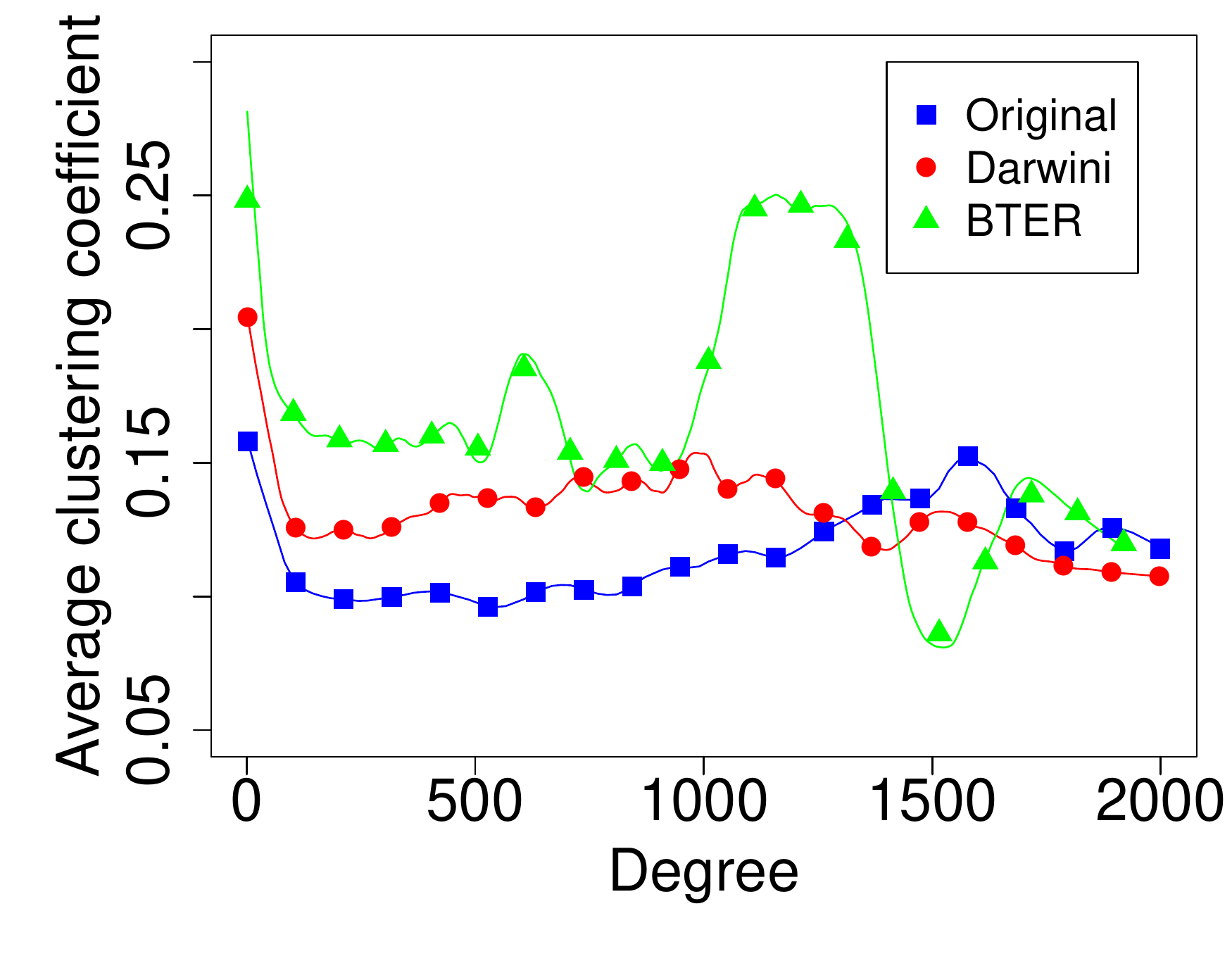}
   \label{fig:twitter_cc}
  }
  \caption{Comparing \sys and BTER on the Twitter graph.}
   \label{fig:twitter}
\end{figure}

Next, we repeat the same experiment on the DBLP co-authorship
graph~\cite{Yang2012a}. Due to
the more manageable size of the DBLP, we were able to fit and generate all the
models described in \cite{Sala:2010:MGM:1772690.1772778} using the publicly
available implementation~\cite{graphmodels}. Here, we evaluate the best
performing models among them, namely Nearest Neighbors~\cite{NN_RW}, Random
Walk~\cite{NN_RW}, dK-2~\cite{Mahadevan:2006} and Forest
Fire~\cite{Leskovec:2005}. 
\begin{table}[t]
 \centering
 \begin{tabular}{| c | p{2cm} | p{2cm} | }
  \hline
  \textbf{Graph} & \textbf{Degree} & \textbf{Clustering Coefficient}\\
  \hline
  \hline
 BTER &  0.21 & 0.64 \\
  \hline
 \sys &  \textbf{0.007} & \textbf{0.19} \\
  \hline
 DK-2 &  \textbf{0.002} & 6.04 \\
  \hline
 Forest Fire &  0.041 & \textbf{0.27}\\
  \hline
 Random Walk &  0.039 & 1.11 \\
  \hline
 Nearest Neighbor & 6.04 & 9.83 \\
  \hline
 \end{tabular}

 \caption{KL-Divergence of degree and clustering coefficient distributions for the DBLP graph.}
 \label{tbl:dblp}
\end{table}

In Figure~\ref{fig:dblp_degree}, we plot the actual distribution and in
Table~\ref{tbl:dblp} we measure the \emph{Kullback-Leibler} (KL) divergence
between the source and the generated distributions for the DBLP graph.
Consistent with the results of \cite{Mahadevan:2006}, dK-2 performs the best
among this set of models.  Nearest Neighbors, one of the best performing models
measured in \cite{Mahadevan:2006}, here tends to produce less low-degree
vertices than expected.  BTER exhibits the same problem, failing to create
high-degree vertices. Notice that \sys exhibits this problem too for this a
graph, but to a lesser extent. Overall, \sys produces the second best degree
distribution among all in terms of the KL-divergence.

We perform the same measurement on the Twitter follower graph~\cite{Kwak2010}.
Here, we compare only \sys and BTER. Figure~\ref{fig:twitter} shows the
results. Both approaches produce a similar degree distribution, though the
produce more high degree nodes than the original distribution. However, \sys
produces a clustering coefficient distribution that is closer to the original
graph than BTER.

\subsubsection{Clustering coefficient distribution}

Here, we use the same graphs as above to compare the accuracy of the generated
clustering coefficient. First, we measure the average clustering coefficient as
a function of the vertex degree for the different models.  We show the result
for the Facebook graph in Figure~\ref{fig:avg_ccd}.

Kronecker underestimates the per degree average clustering coefficient by up to
4 orders of magnitude. BTER performs better than Kronecker as it by design
attempts to produce a graph with a high average clustering coefficient. Even
so, notice the clustering coefficient diverges significantly for high-degree
nodes. Specifically, for nodes with degree higher than 2500, the clustering
coefficient could by off by an order of magnitude. Again, BTER cannot produce
vertices with high degrees.  Instead, for \sys the average clustering
coefficient differs follow closely the source distribution across the entire
spectrum of degrees.

Figure~\ref{fig:dblp_cc} compares the per degree average clustering coefficient
between \sys and the rest of the models on the DBLP graph. While in terms of
degree distribution the other models produced good results, most of the models
underestimate the average clustering coefficient by at least X\%. Only BTER can
capture the average clustering coefficient. Still, \sys outperforms BTER
especially for high-degree vertices.  Interestingly, the source DBLP graph
exhibits an increase in the clustering coefficient for vertices with degrees
between 100 and 160. Both \sys and BTER are able to reproduce this artifact.

Further, for the Facebook graph, we also measure the distribution of the
clustering coefficient values across the entire graph.  We show this result in
Figure~\ref{fig:ccd}.  the clustering coefficient distribution. As expected,
Kronecker produces only vertices with low clustering coefficient. BTER tends to
produce many vertices with high clustering coefficient. \sys captures the
actual source distribution better than all models.

\subsubsection{Joint degree distribution}

\begin{figure*}[ht]
  \subfigure[Degree 5] {
      \includegraphics[width=0.31\linewidth]{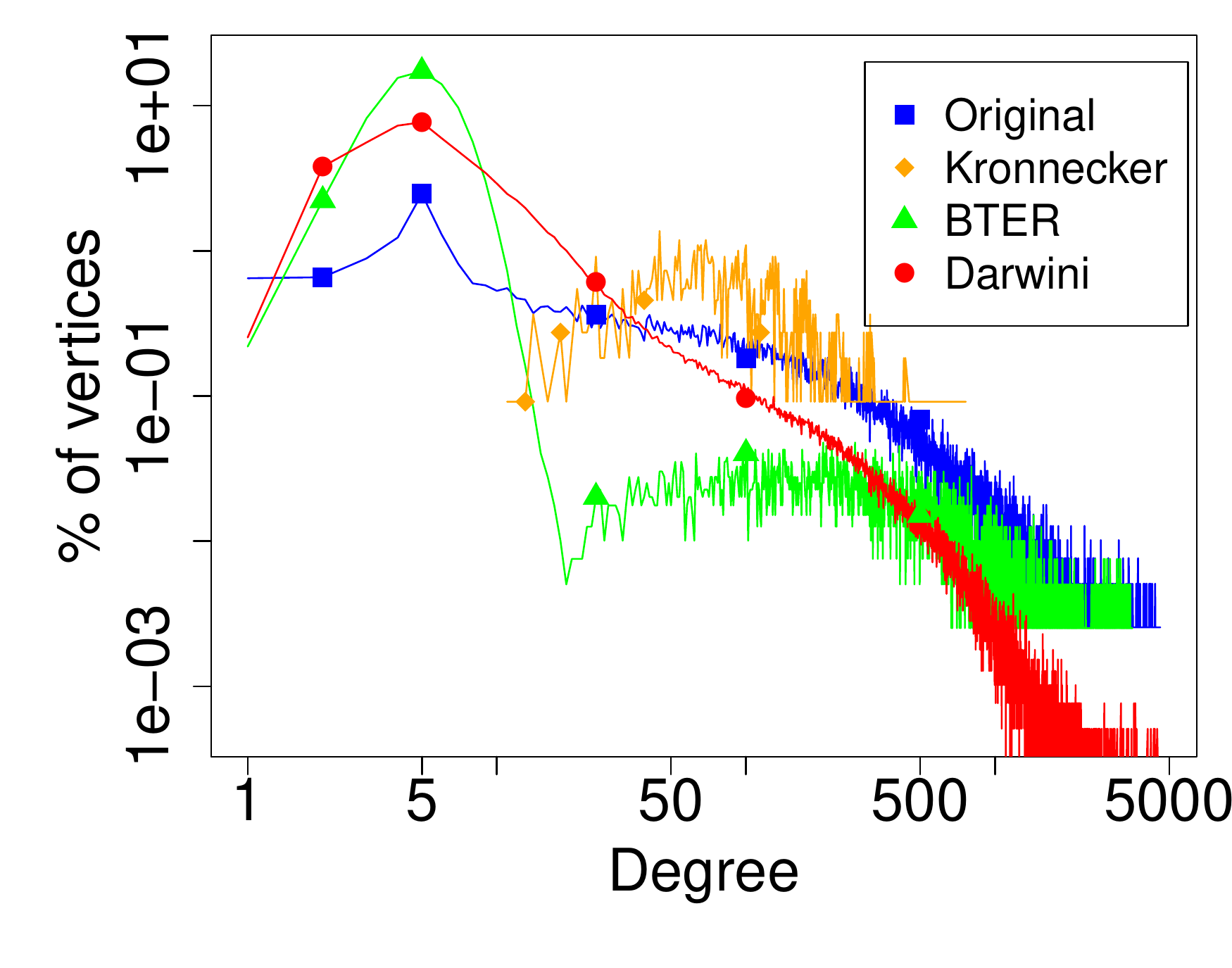}
      \label{fig:join_degree_5}
  }
  \subfigure[Degree 32] {
    \includegraphics[width=0.31\linewidth]{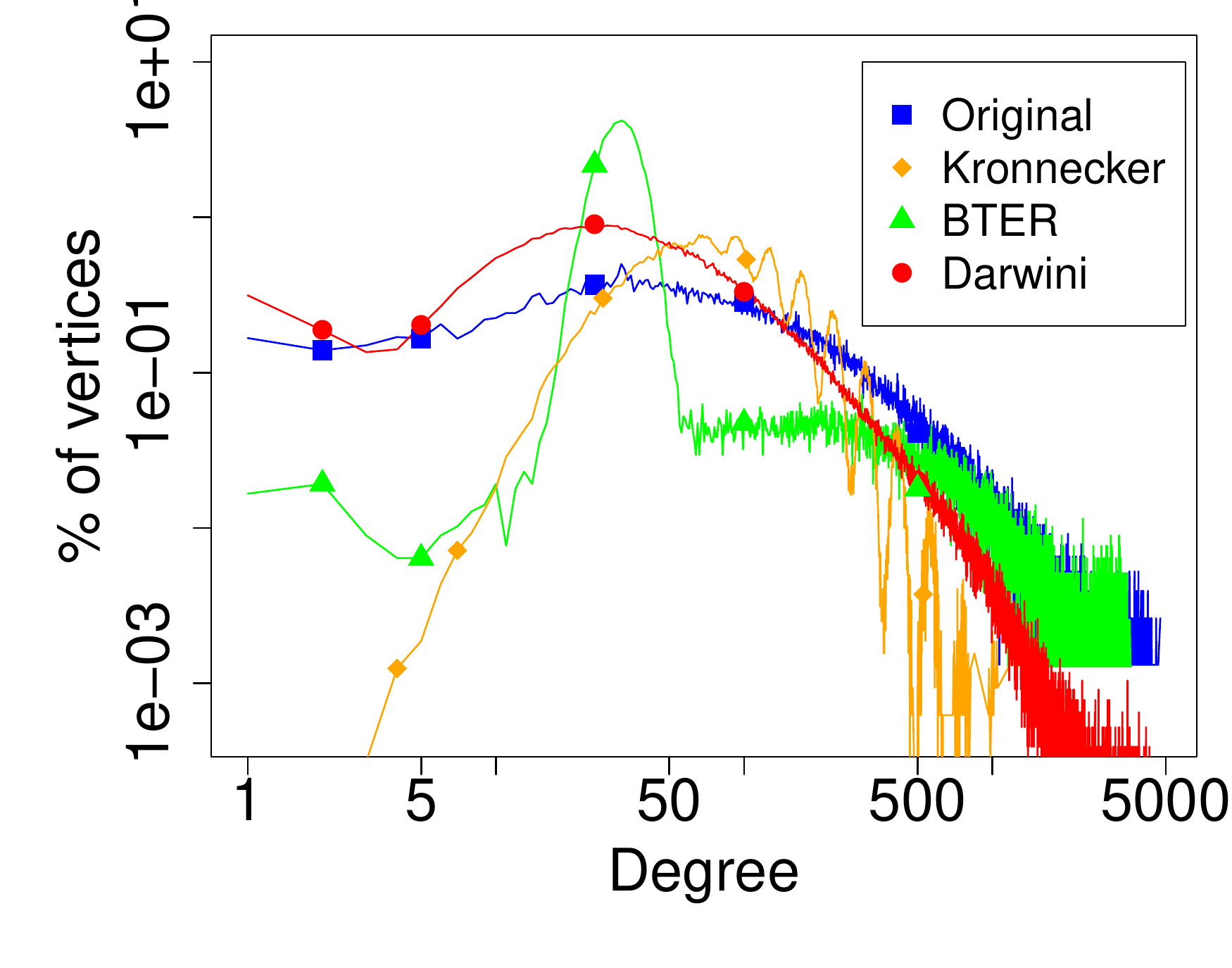}
    \label{fig:join_degree_32}
  }
  \subfigure[Degree 500] {
    \includegraphics[width=0.31\linewidth]{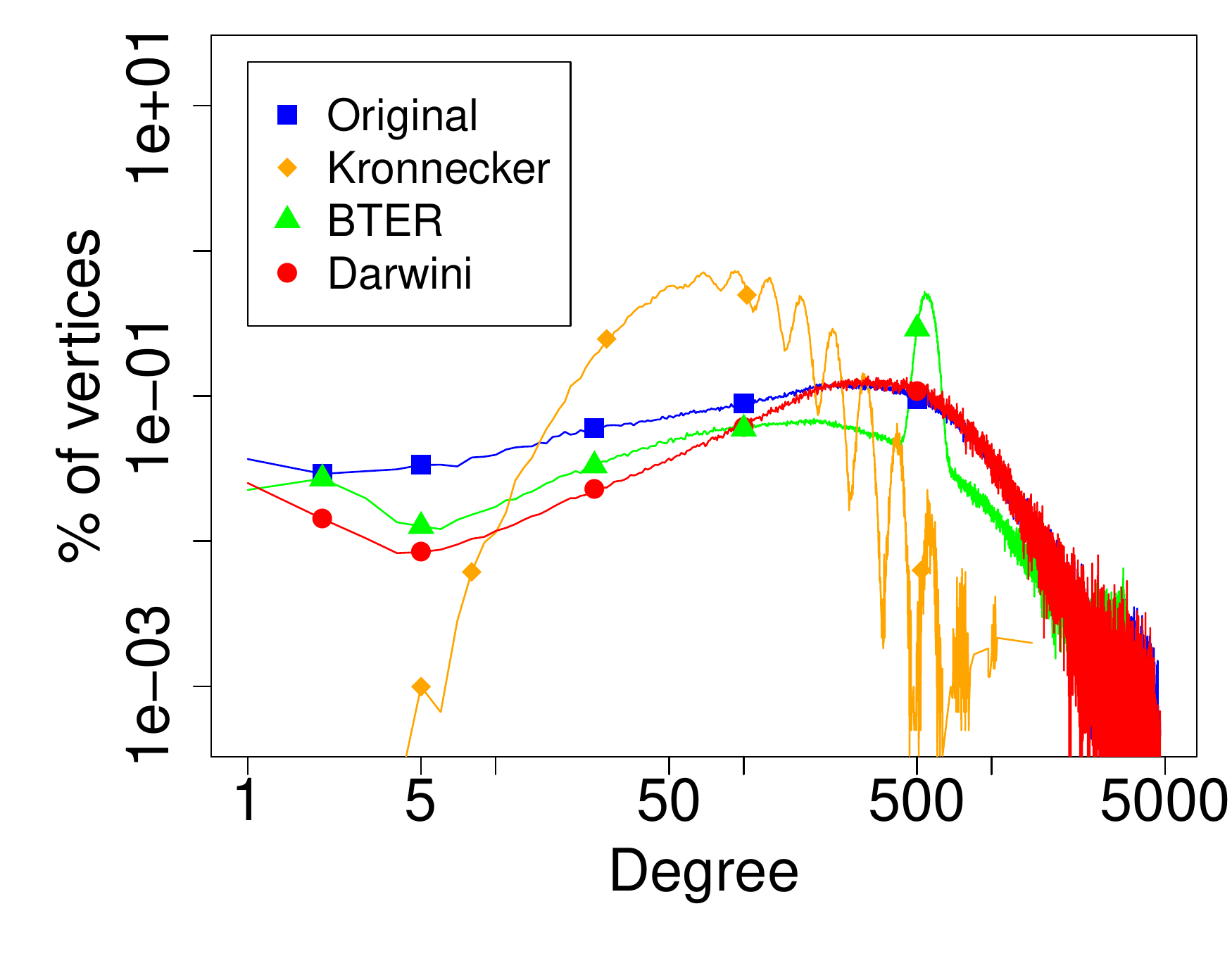}
    \label{fig:join_degree_500}
  }
  \caption{Joint degree distribution on the source Facebook graph and the graphs generated by \sys, BTER and Kronecker.}
  \label{fig:join_distr}
\end{figure*}

\sys tries to produce a realistic joint-degree distribution. Here, we measure
how close to the original Facebook graph the generated joint-degree
distribution is for \sys, BTER and Kronecker.  In Figure \ref{fig:join_distr},
we demonstrate the joint-degree distribution for vertices with degree 5, 32 and
500.

First notice that the distribution produced by Kronecker diverges the most from
the original one.  The BTER model improves upon Kronecker, but still produces a
skewed joint degree distribution. This is due to grouping only vertices with
the same degree into the same block. As a result, more vertices with same
degree are connected to each other than in the original graph.  Instead, by
grouping vertices into the bucket based on the value of the $c_id_i(d_i-1)$ product,
\sys allows the connection of more diverse vertices with respect to degree.

\begin{table}[t]
 \center
 \begin{tabular}{| c | c | c | c |}
  \hline
  \textbf{Distribution} & \textbf{Kronecker} & \textbf{BTER} & \textbf{\sys} \\
  \hline
  \hline
   Degree & 3.82 & 0.02 & 0.0014 \\
  \hline
   Joint Degree, d=5 & N/A & 0.57 & 0.11 \\
  \hline
   Joint Degree, d=32 & 0.48 & 0.27 & 0.17 \\
  \hline
   Joint Degree, d=500 & 1.56 & 0.34 & 0.012 \\
  \hline

 \end{tabular}

 \caption{KL-divergence between the original Facebook graph and the generated
graph distributions.} 
  \label{tbl:kl_divergence}
\end{table}

We also measured the KL-divergence of the joint-degree distributions between
the original graph and generated graphs. The result, shown in
Table~\ref{tbl:kl_divergence}, verifies that \sys produces a more accurate
distribution. Notice that for degree $d=5$, we cannot estimate the
KL-divergence for the Kronecker model as it does not produce enough vertices
with this degree.

\subsubsection{PageRank distribution and eigenvalues}

\begin{figure*}[ht]
  \subfigure[PageRank] {
      \includegraphics[width=0.31\linewidth]{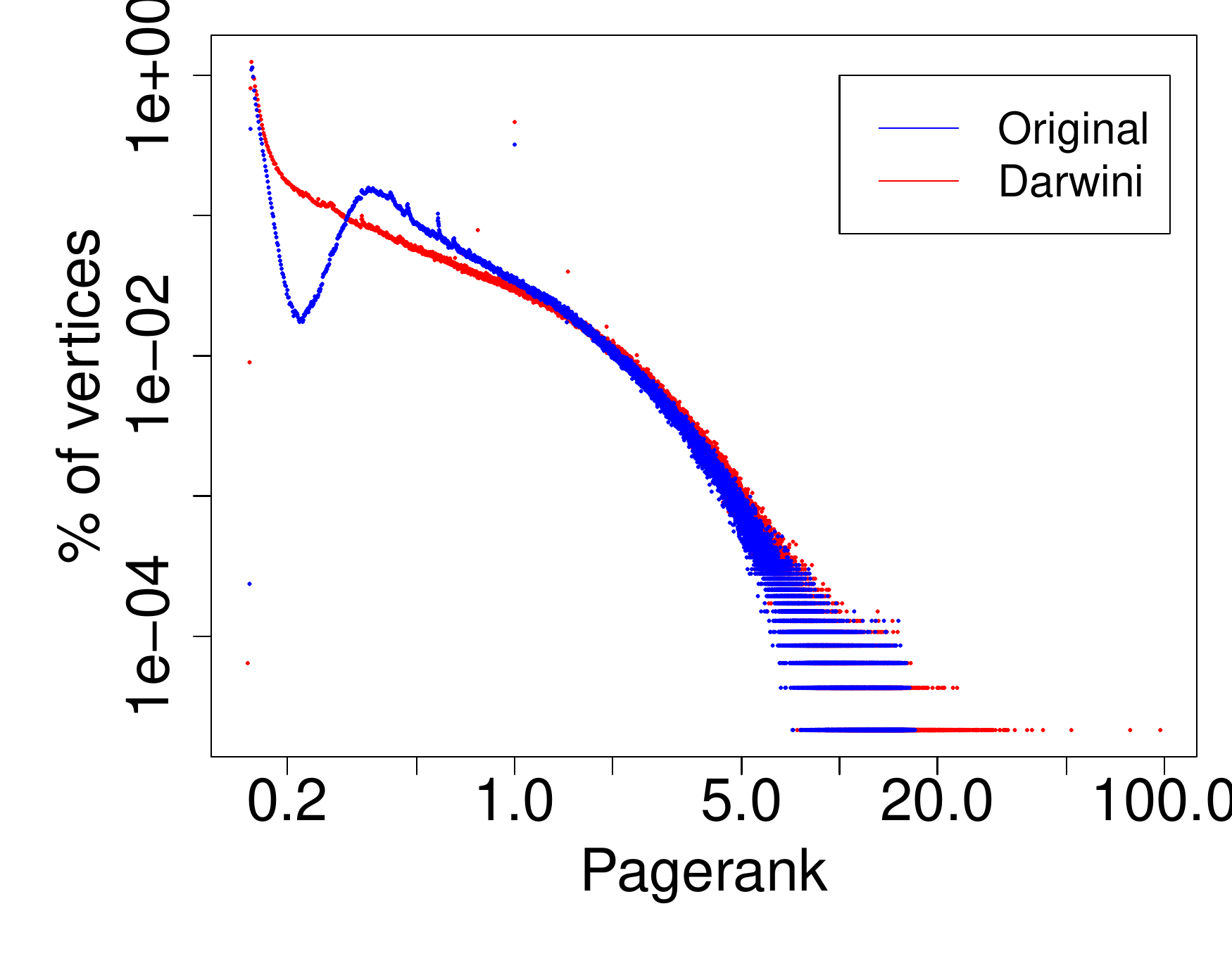}
      \label{fig:pr1}
  }
  \subfigure[PageRank] {
    \includegraphics[width=0.31\linewidth]{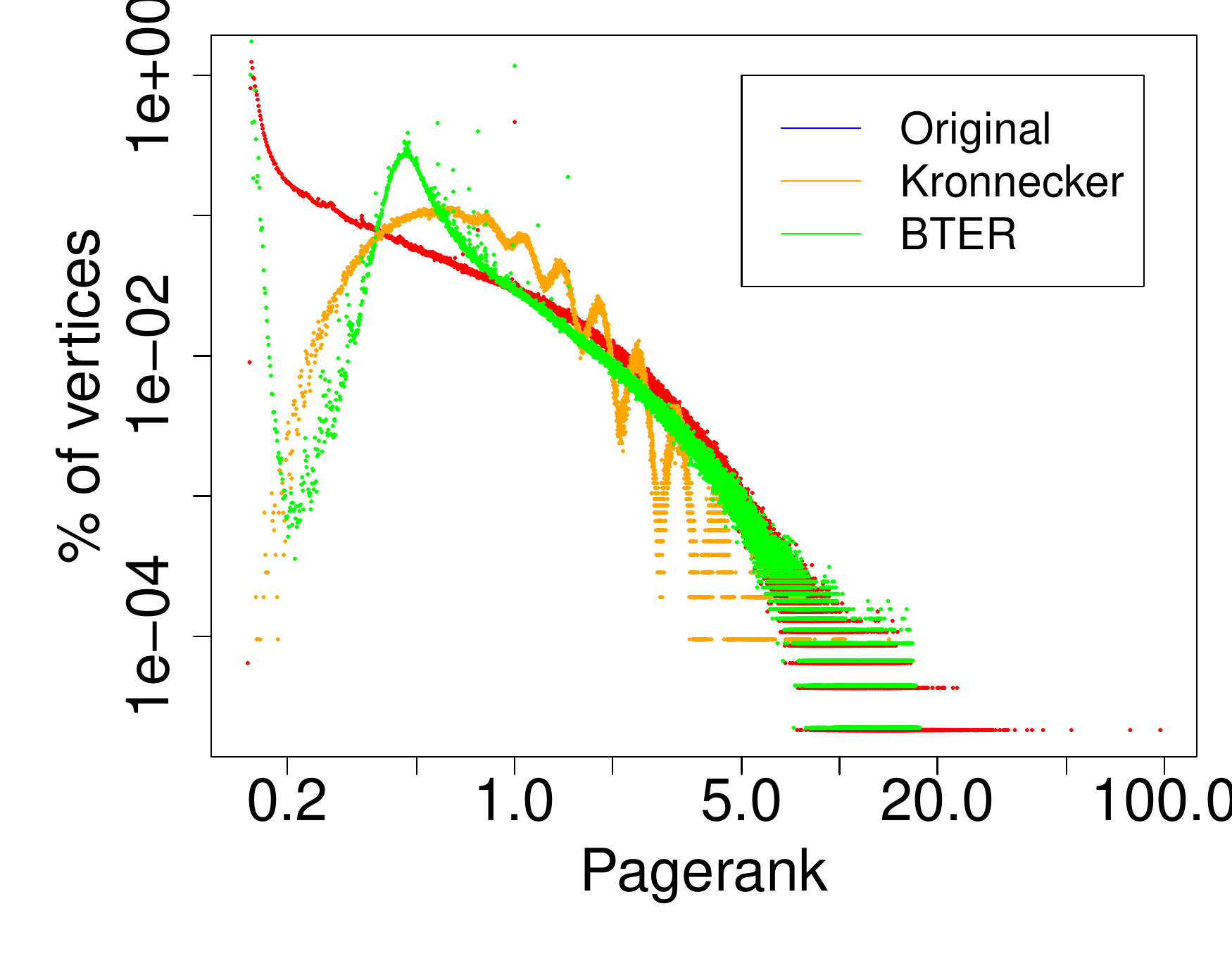}
    \label{fig:pr2}
  }
  \subfigure[Eigenvalues] {
    \includegraphics[width=0.31\linewidth]{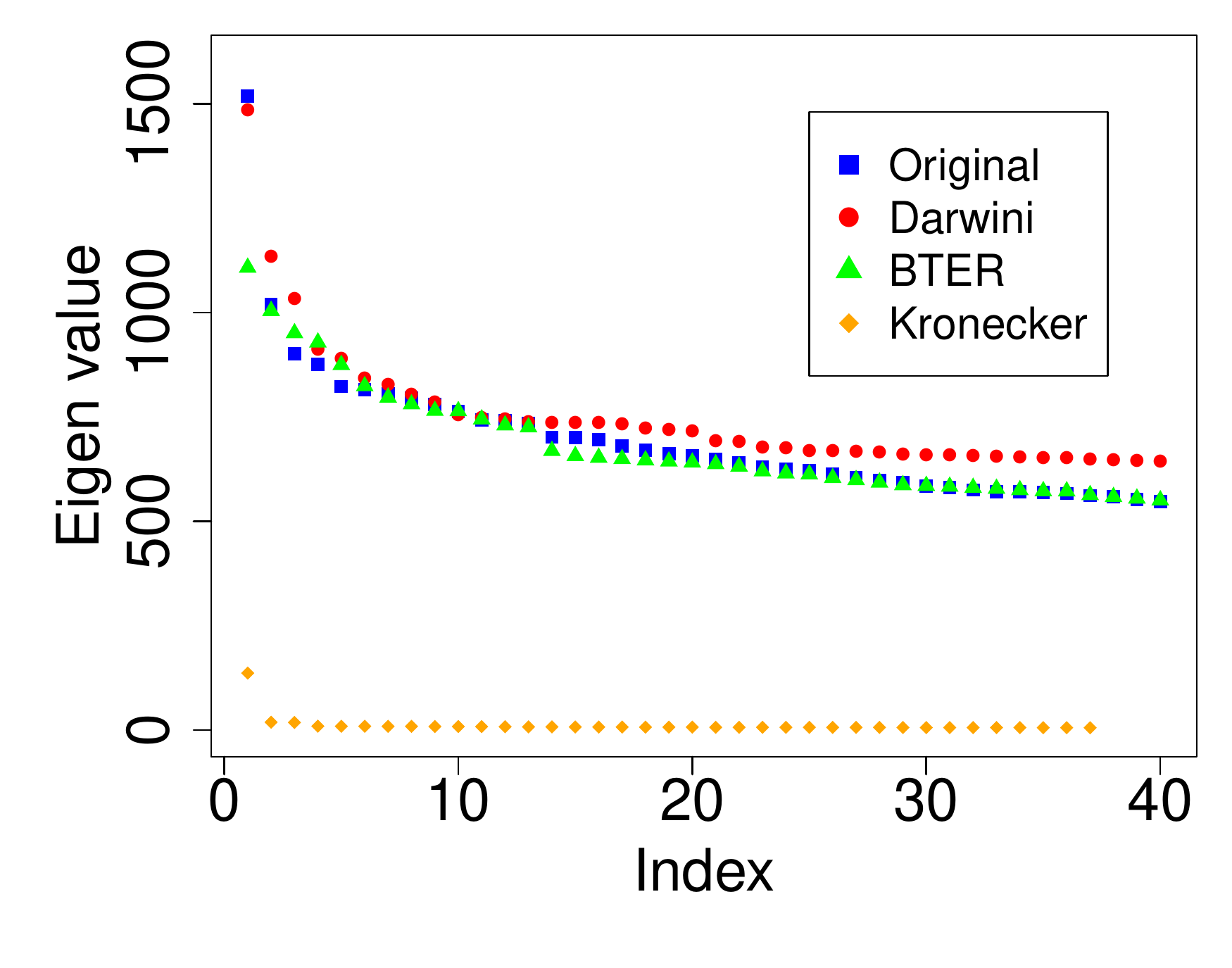}
    \label{fig:eigen}
  }
  \caption{PageRank and Eigenvalues on the source Facebook graph and the graphs generated by Darwini, BTER and Kronecker.}
  \label{fig:pr_eigen}
\end{figure*}

The PageRank distribution and graph eigenvalues are common metrics used to
characterize a graph structure. In Figures \ref{fig:pr1} and \ref{fig:pr2}, we
compare the PageRank distributions between \sys, BTER and Kronecker, while
Figure~\ref{fig:eigen} shows the eigenvalues of the original and the generated
graphs.

Although graphs generated by \sys exhibit better PageRank distributions than
other models, notice that the distribution has a significant dip caused by the
block structure created at the initial stage. We hypothesize that this is due
to the fact that real graphs have more hierarchical and overlapping community
structure, while \sys strictly assigns every vertex to one community.  Further,
notice that both BTER and \sys generate graphs with similar distribution of
eigenvalues. \sys tends to overestimate the values at the tail of the eigenvalue
spectrum.

\subsubsection{K-Core decomposition}

\begin{figure*}[ht]
  \centering
  \subfigure[Original] {
   \includegraphics[width=0.23\linewidth]{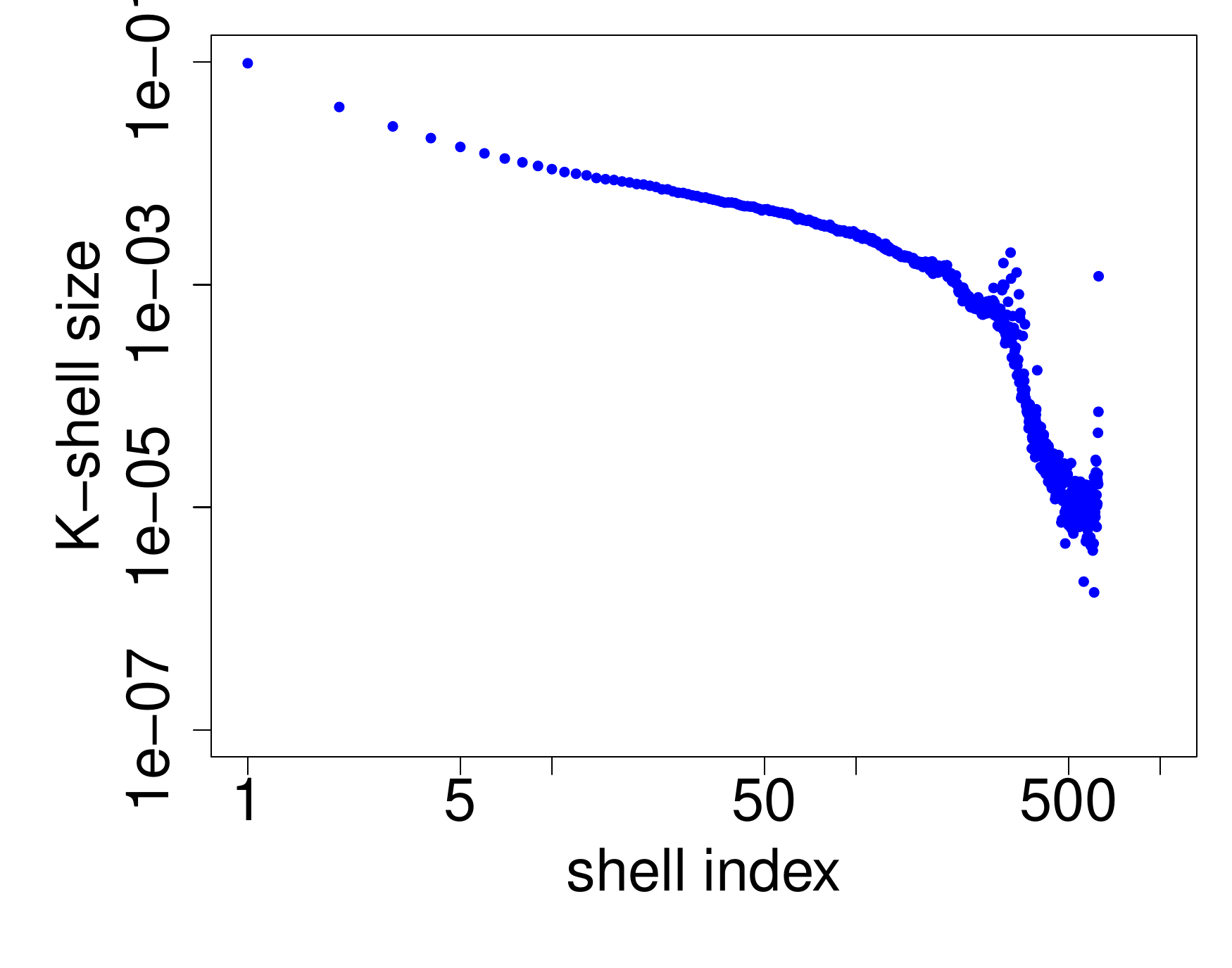}
   \label{fig:kcore_original}
  }
  \subfigure[\sys] {
   \includegraphics[width=0.23\linewidth]{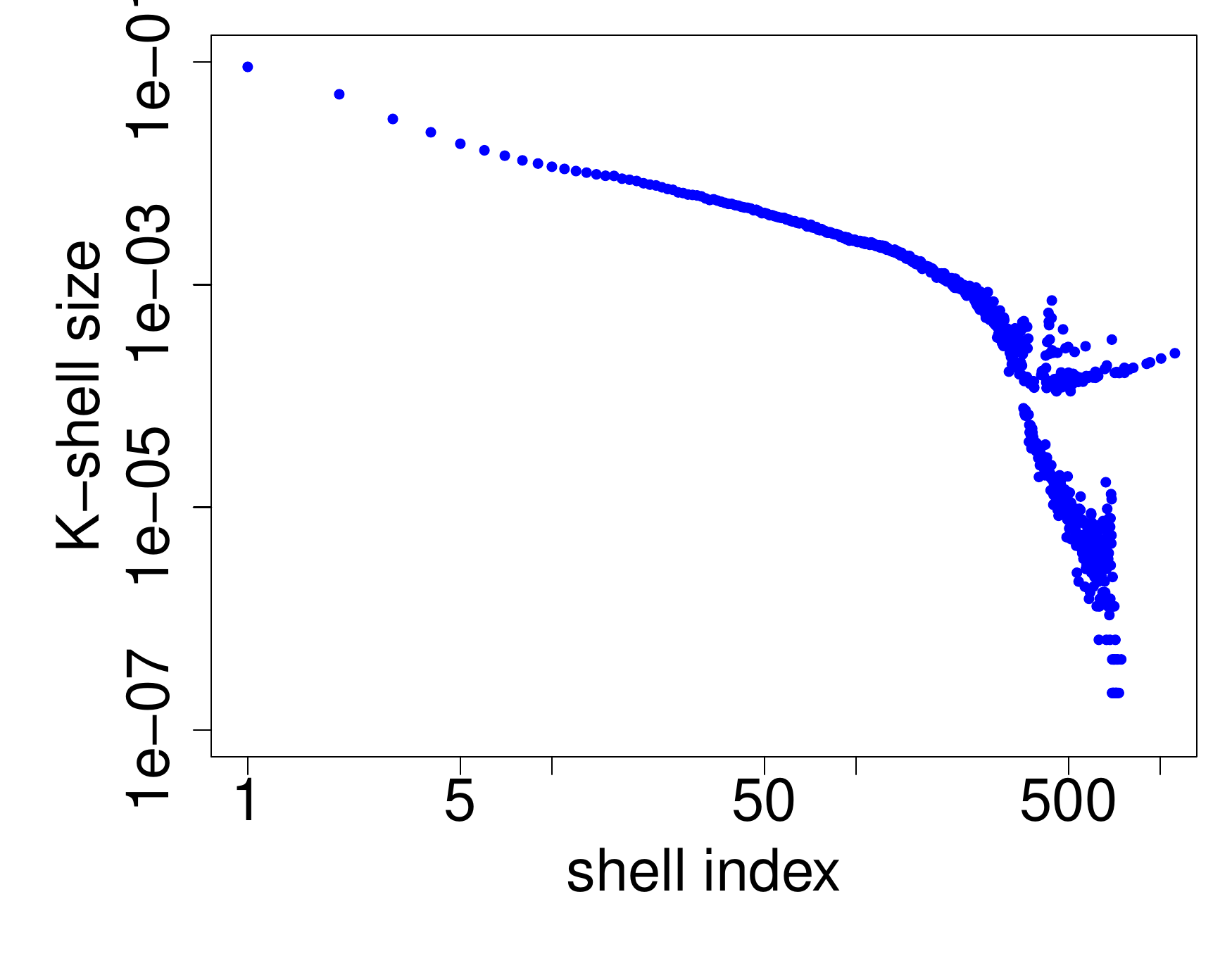}
   \label{fig:kcore_ours}
  }
  \subfigure[BTER] {
   \includegraphics[width=0.24\linewidth]{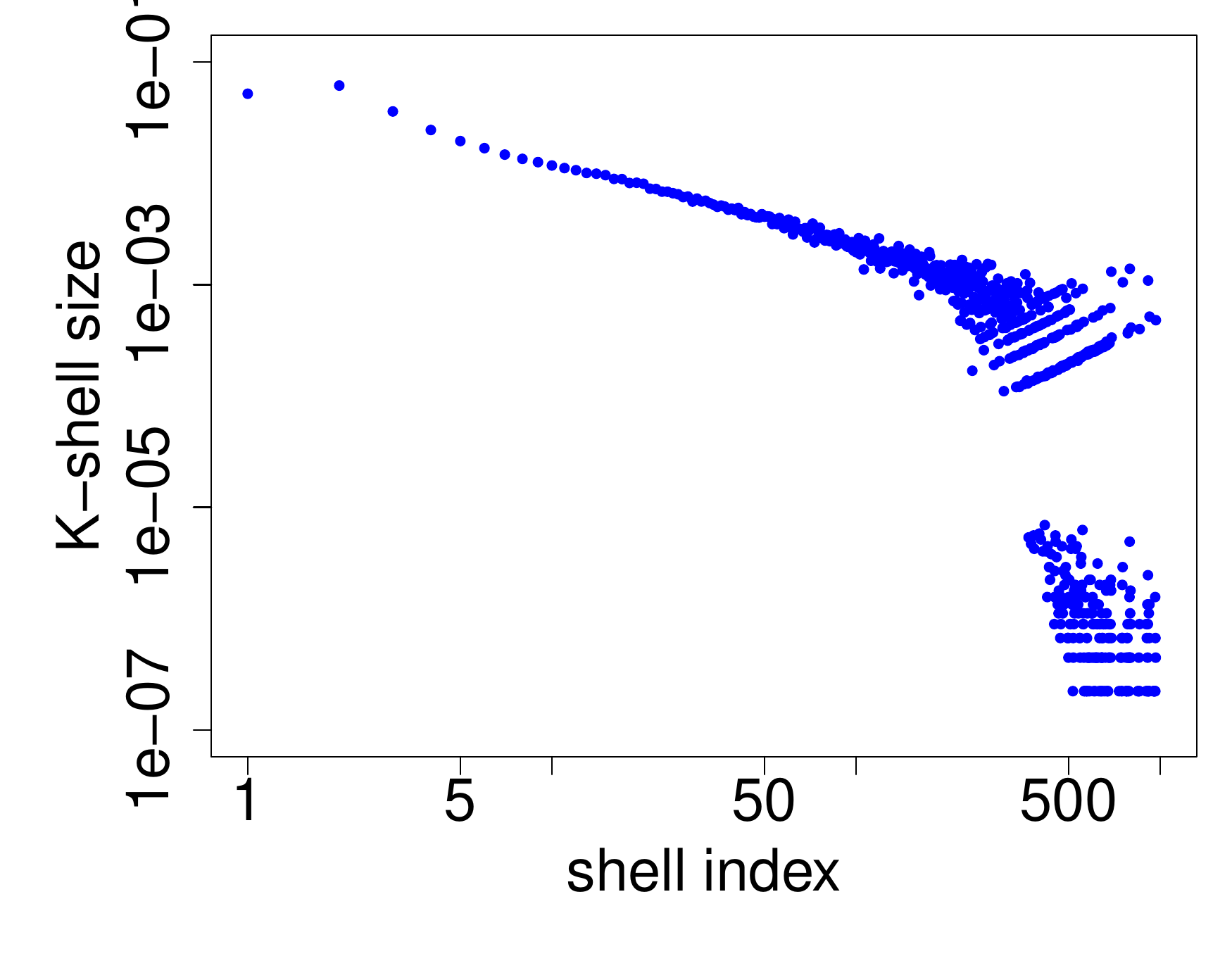}
   \label{fig:kcore_bter}
  }
  \subfigure[Kronecker] {
   \includegraphics[width=0.23\linewidth]{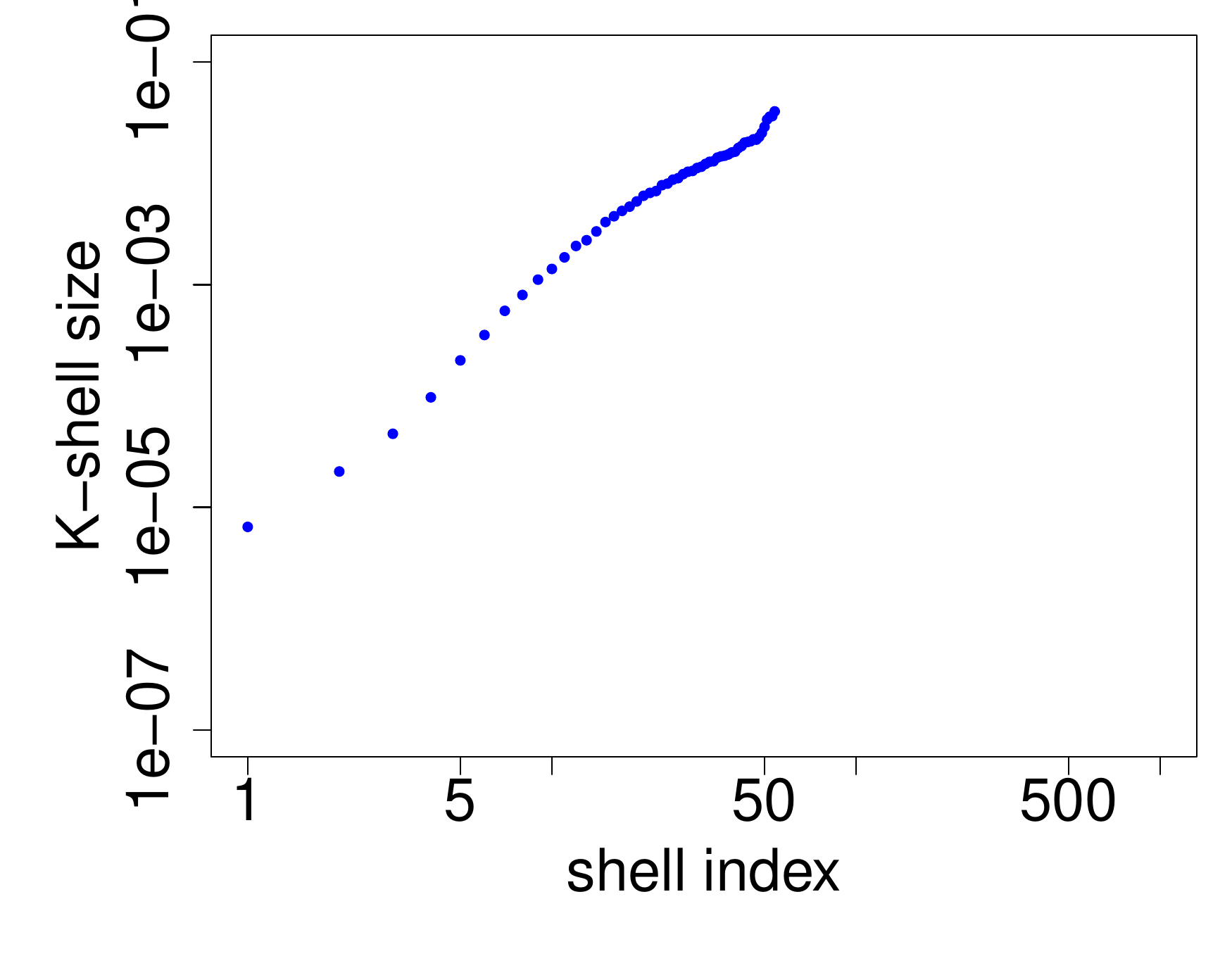}
   \label{fig:kcore_kronecker}
  }
  \caption{K-Core decomposition on the source Facebook graph and the graphs generated by Darwini, BTER and Kronecker.}
   \label{fig:kcore}
\end{figure*}

The K-core decomposition of a graph is typically used to study hierarchical
properties of a graph such as finding regions of high centrality and
connectedness.  The K-Core decomposition is computed by recursively eliminating
weakly connected vertices, and is measured by the size of the shells obtained
through this recursive elimination~\cite{DBLP:journals/corr/abs-cs-0511007}.
In Figure~\ref{fig:kcore}, we plot the shell sizes of the original and the
generated graphs.

The K-core decomposition of \sys is the closest to that of the original graph.
The difference in shell size for high shell indexes can be attributed to the
block structure, in particular, the fact that each vertex belongs to a single
block, while in the real graph vertices belong to multiple hierarchical and
overlapping communities.

\subsubsection{Connected components}

Real graphs usually contain a giant connected component and a number of small
components. Here, we evaluate the ability of \sys to capture this property, and
compare with BTER and Kronecker. Table~\ref{tbl:connected_components} shows the
number of components and the size of the giant component as a percentage of the
total graph size. \sys produces a giant components with a similar size and a
set of small components. This holds true for BTER as well, while the Kronecker
model tends to produce 1 or 2 connected components.

\begin{table}[t]
 \center
 \begin{tabular}{| c | c | c |}
  \hline
  \textbf{Generator} & \textbf{\#components} & \textbf{Size of GC}  \\
  \hline
  \hline
   Original graph & 11K & 99.33\% \\
  \hline
   Kronecker graph & 2 & 99.86\% \\
  \hline
   BTER & 27K & 98.1\% \\
  \hline
   \sys & 2.3K & 99.83\% \\
  \hline
 \end{tabular}

 \caption{Number of connected components and the relative size of the giant connected component in the original and the generated graphs.}
 \label{tbl:connected_components}
\end{table}

\subsection{Impact on applications}

One of our initial motivations was to use \sys to allow researchers to
benchmark graph processing systems on a reference graph, for instance the
Facebook social graph, without sharing the graph. Here, we measure how
representative the synthetic graphs are in terms of the system performance.

In this experiment, we use as source a Facebook connected subgraph with 300M
vertices.  Using this source, we generate synthetic graphs with \sys, BTER and
Kronecker.  Subsequently, we run a variety of graph mining application,
developed on the Apache Giraph framework, on all these graphs and compare the
observed performance of the Apache Giraph system. Here, we run four different
applications: PageRank, Clustering Coefficient, Eigenvalue decomposition and
Balanced Partitioning~\cite{socialhash}.

\begin{figure}[h]
  \includegraphics[width=1\linewidth]{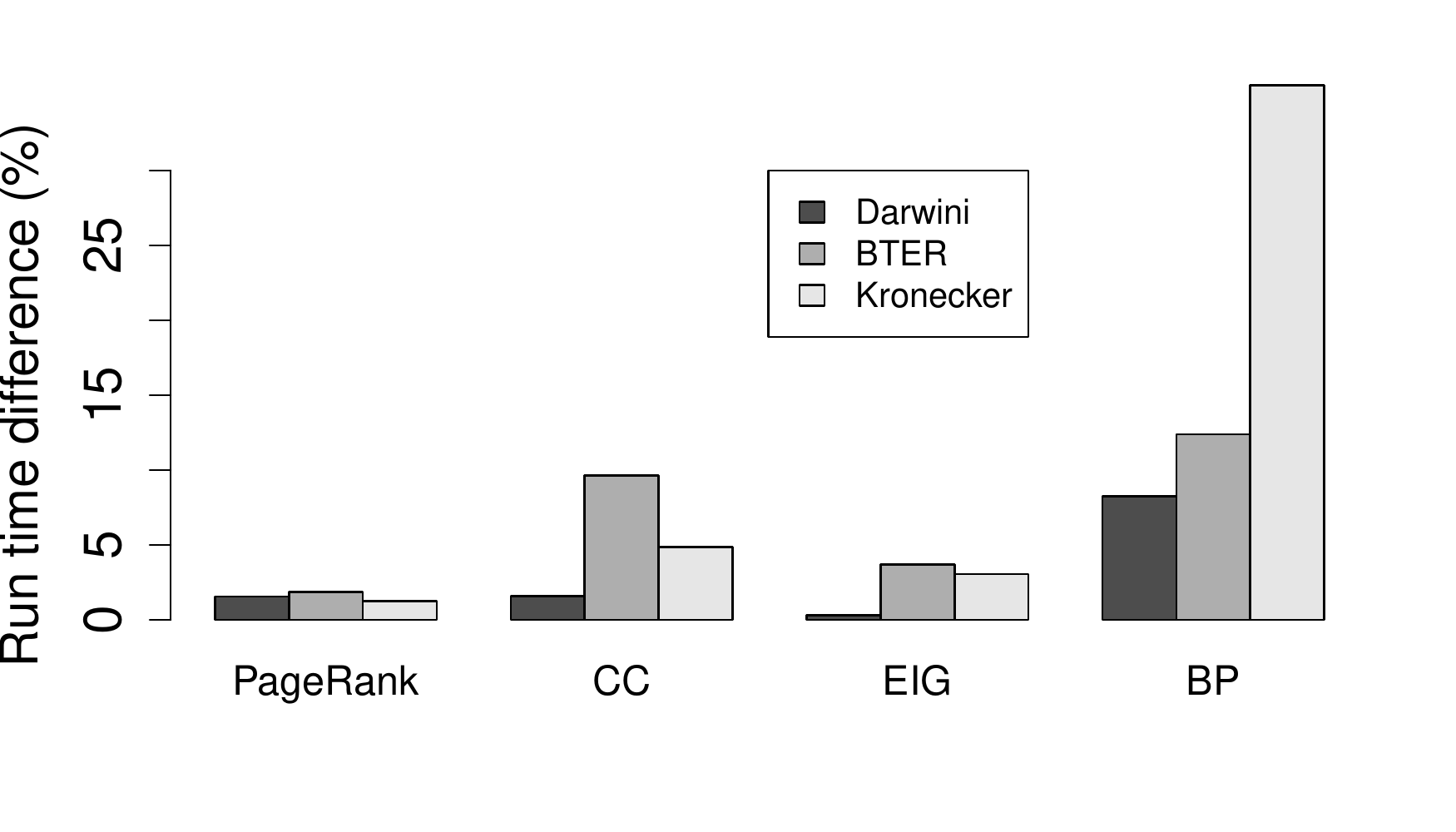}
  \caption{Impact of graph structure on system performance. The graph shows the
relative performance difference compared to processing the original graph.}
  \label{fig:impact}
\end{figure}

Figure~\ref{fig:impact} shows the relative difference in runtime between the
original and the synthetic graphs for the different applications. Each data
point is an average of three runs. First, notice that for PageRank the
difference is small for all graphs. The computation overhead of this
application is proportional to the number of edges in the graph.  Giraph
distributes the graph across machines randomly, therefore, the same applies for
the incurred network overhead. Since all synthetic graphs have almost the same
number of edges with the original graph.

The difference in performance becomes more apparent for the rest of the
applications because of their computation and communication patterns. For
instance, in the clustering coefficient vertex-centric algorithm, every vertex
creates a message that is proportional in size to its degree and sends it to
all its neighbors. Even though the number of edges is the same in all graphs, a
different clustering can impact the size of the messages and, hence, the
observed application performance. In these case, the observed performance on
the graph generated with \sys is closer to the one on the original graph.

\subsection{Scalability}

\begin{figure}[h]
  \centering
  \subfigure[Time vs. graph size] {
  \includegraphics[width=0.48\linewidth]{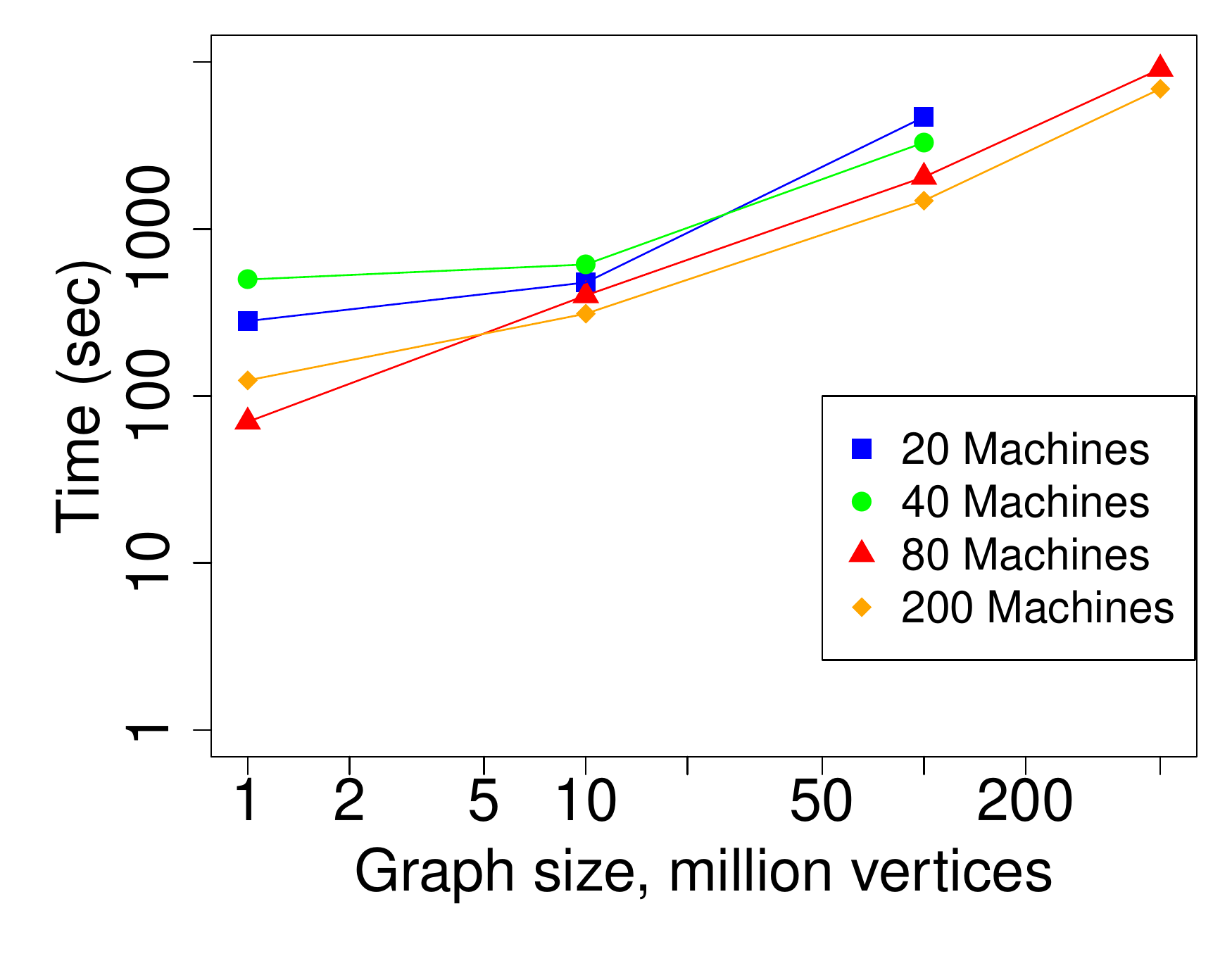}
  \label{fig:scalability1}
  }
  \hspace{-1.7em}
  \subfigure[Time vs. \#machines] {
  \includegraphics[width=0.48\linewidth]{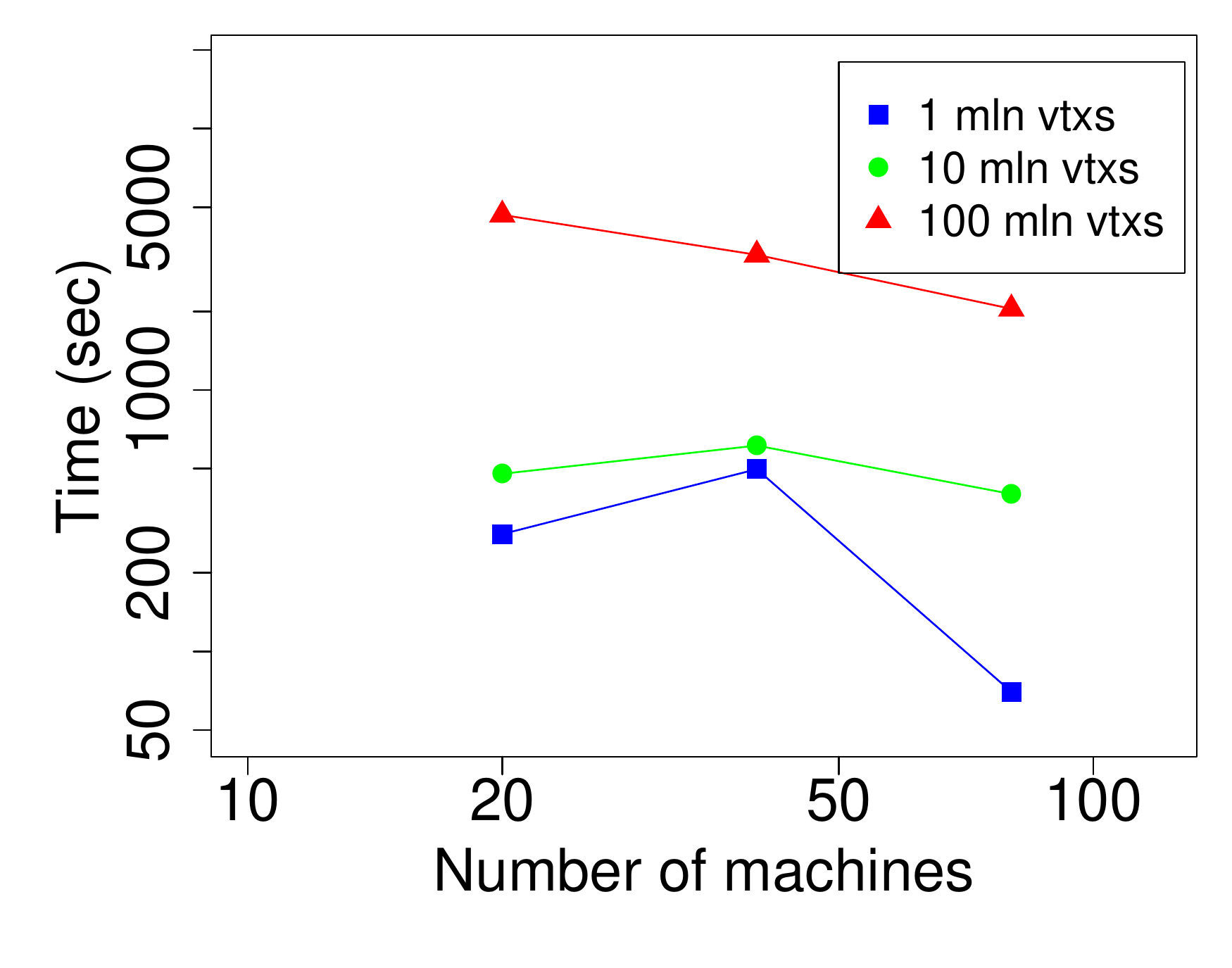}
  \label{fig:scalability2}
  }
  \caption{(a) \sys scales linearly on the number of vertices. (b)
For large graphs, \sys scales linearly as the compute cluster size increases.}
  \label{fig:scalability}
\end{figure}

Here, we evaluate the scalability of the \sys implementation.  We use an
experimental cluster with 200 machines, each with 256G of RAM and 48 cores.
Figure~\ref{fig:scalability1} show the time to generate a graph as a function
of the output graph size. The graph generation time scales linearly with the
number of vertices until we hit the memory limit.  In
Figure~\ref{fig:scalability2}, we show how the graph generation time improves
as we increase the size of the compute cluster. For a sufficiently large graph,
the time decreases linearly.  Smaller graph sizes do not benefit from a large
number of machines due to the network overhead.

Further, we used \sys to generate a scaled-up version of the Facebook social
graph. We used the entire Facebook social graph as the source graph and
generated a synthetic graph with one trillion edges. This task took
approximately 7 hours on the same 200-machine compute cluster Although we omit
the details, the generated distributions are close to the source distribution,
consistent with our result on the smaller subgraph.

\section{Related work} 
\label{sec:related}
In this section, we briefly introduce several well-known social graph
generation models.

Our work is inspired by the Block Two-Level Erd\"os-R\'enyi (BTER)
\cite{DBLP:journals/corr/abs-1302-6636, DBLP:journals/corr/abs-1112-3644}
model. As we show in our evaluation, the BTER model is capable of capturing the
average clustering coefficient, but fails in generating high-degree vertices
and often results in graphs with skewed joint degree distribution.

The Barabasi-Albert model~\cite{Barabasi509} uses the  preferential attachment
mechanism to produce random graphs with power-law degree distributions.
However, preferential attachment does not generally produce higher than random
number of triangles, resulting in graphs with low clustering coefficient.

The Random Walk model \cite{NN_RW} simulates the randomized walk behavior of
friend connections in a social network.  Each node performs a random walk
starting from a randomly chosen node in the graph, and randomly connects to a
new node with a given probability.  The Nearest Neighbor model \cite{NN_RW} is
based on idea that people sharing a common friend are more likely to become
friends. Therefore, graph generation goes as follows: after a new node is
connected to an existing node, random pairs of the 2-hop neighbors are also
connected with specified probability.  While Random Walk and Nearest Neighbor
models are relatively accurate in terms of degree distribution and clustering
coefficient, they are biased towards inter-connecting high-degree nodes, and
produce graphs with significantly shorter path lengths and network
diameter~\cite{Sala:2010:MGM:1772690.1772778}.

Kronecker graphs \cite{Leskovec:2010:KGA:1756006.1756039}  are generated by
recursive application of Kronecker multiplication to an initiator matrix.  The
initiator matrix is selected by applying the KronFit algorithm to the original
graph. Modifying the size of the initiator matrix introduces a tradeoff between
overhead and accuracy. Generally, increasing the size of initiator matrix
results in better accuracy, but increases the fitting time. In our
experimentation, we found it hard to apply the existing KronFit implementation
to real size graphs.

DK-graphs \cite{Mahadevan:2006} is a family of stochastically generated graphs
that match the respective DK-series of original graph. DK-1 graphs match the
degree distribution of the original graph, while DK-2 matches the joint degree
distribution.  DK-3 matches the corresponding DK-3 series, including the
clustering coefficient of the original graph.  However generating DK-3 graph
using rewiring incurs very high overhead. We are not aware of any efficient
algorithm that generates large DK-3 graphs.

The LDBC Social Network Benchmark\cite{SocialNetworkBencmark} is based on the
idea of emulating user profiles and behaviors. Although very powerful, this
approach requires to specify many parameters that are hard to fit. It is also
limited to friendship graphs, while we must generally be able apply this on
different types of entities and relationships.

\section{Conclusion and future work}
\label{sec:concl}
This paper introduced Darwini, a scalable synthetic graph generator that can
accurately capture important metrics of social graphs, such as degree,
clustering coefficient and joint-degree distributions.  We implemented \sys on
top of a graph processing framework, making it possible to use it on any
commodity cluster.  Even so, to facilitate access to large-scale datasets,
apart from open sourcing \sys, we also intend to make generated graph datasets
publicly available as well.

At the same time, we believe there are interesting future directions in this
area.  For instance, real social network users typically belong to multilpe
communities, based on workplace, university affiliation, and others, affecting
the connectivity of the graph. However, \sys and other models assign vertices
to a single community.  Capturing the multi-community structure will provide
more accurate synthetic datasets.  Furthermore, current generators focus on the
graph structure, and lack models for generating metadata, such as community
labels characterizing vertices, or user similarity metrics characterizing
edges.  Such data will enable research in a variety of areas such as community
detection algorithm, without the need to share the original data and
potentially compromize user privacy.

\bibliographystyle{shortabbrv}
\bibliography{graphs,nondoc}  

%
%
\balancecolumns

\end{document}